\documentclass[preprint]{aa}
\usepackage{txfonts}
\usepackage{graphicx,subfigure}
\usepackage{epsfig}
\usepackage{epsf}
\usepackage{lscape}
\usepackage{natbib}
\usepackage{rotating}
\usepackage{tabularx}
\newcommand{\mic}{\,{\rm \mu m} }

\begin{document}

\title{Revisiting the dust properties in the molecular clouds of the LMC}

\author{D. Paradis \inst{1,2} 
  \and
  C. Mény \inst{1,2}
  \and M. Juvela \inst{3}
  \and A. Noriega-Crespo \inst{4}
  \and I. Ristorcelli \inst{1,2}
  }  
\institute{CNRS; IRAP; 9 Av. du Colonel Roche, BP 44346, F-31028, Toulouse, cedex
4, France 
\and
Universit\'e de Toulouse; UPS-OMP; IRAP; Toulouse, France 
\and
Department of Physics, P.O.Box 64, FI-00014, University of Helsinki, Finland
\and
Space Telescope Science Institute, 3700 San Martin Drive, Baltimore,
MD 21218, USA
}
\authorrunning{Paradis et al.}
\titlerunning{}
\date{}
\abstract
{Some Galactic molecular clouds show signs of dust evolution as
  compared to the diffuse interstellar medium, most of the time
  through indirect evidence such as color ratios, increased dust
  emissivity, or scattering (coreshine). These signs are not a feature
  of all Galactic clouds. Moreover, molecular clouds in the Large Magellanic Cloud (LMC) have been
  analyzed in a previous study based on Spitzer and IRIS data, at
  4$^{\prime}$ angular resolution, with the use of one single dust
  model, and did not show any signs of dust
  evolution.} 
{In this present analysis we investigate the dust properties
  associated with the different gas phases (including the ionized phase
  this time) of the LMC molecular clouds at
  1$^{\prime}$ angular resolution (four times greater than the previous
  analysis) and with a larger spectral coverage range thanks to
  Herschel data. We also
 ensure the robustness of our results in the framework of various dust models.}
{We performed a decomposition of the dust emission in the infrared (from
  3.6 $\mic$ to 500 $\mic$) associated with the atomic, molecular, and
  ionized gas phases in the molecular clouds of the LMC. The resulting spectral
  energy distributions were fitted with four distinct dust models. We then analyzed the model
  parameters such as the intensity of the radiation field and the
  relative dust abundances, as well as the slope of the emission
  spectra at long wavelengths. }  
{This work allows dust
  models to be compared with infrared data in various environments for the first time, which reveals
  important differences between the models at short wavelengths in
  terms of data fitting 
(mainly in the polycyclic aromatic hydrocarbon bands). In addition, this analysis points out distinct
results according to the gas phases, such as dust composition
directly affecting the dust temperature and the dust emissivity
in the submillimeter (submm), and different dust emission in the near-infrared (NIR).}
{We observe  direct evidence of dust property evolution from the diffuse to the
  dense medium in a large sample of molecular clouds in the LMC. In addition, the differences in the dust component abundances
  between the gas phases could indicate different origins of grain
  formation. We also point out the presence of a NIR-continuum in all
gas phases, with an enhancement in the ionized gas. We favor the hypothesis of
an additional dust component as the carrier of this continuum.   }


\maketitle
\section{Introduction}
The study of molecular clouds is important to understand the process
of star formation. Whereas some clouds show activity of
high-mass star formation, others seem to be quiescent and preferentially lead
to low-mass star formation or to no star formation at all \citep{Lis01}. It is therefore assumed that young
stars and HII regions are born inside the molecular clouds. Ionized gas
is then often located in the vicinity of the clouds. We therefore expect to observe
changes in the properties of dust associated with the molecular clouds, and especially
in the different phases of the gas inside and surrounding the cloud. Indeed, the dense molecular phase
could be shielded from the interstellar radiation field and ensure the
survival of small grains and/or molecules in this phase. In addition, ice
mantles could cover the largest grain surfaces and allow dust
aggregation processes. However, aggregation is not directly
observable but could be linked to the decrease of the IRAS 60/100
$\mic$ observed in some clouds \citep{Laureijs91}, potentially resulting from the
disappearance of the small grains. Changes in the dust
emissivity have been observed in some molecular clouds
\citep{Stepnik03, Paradis09} and
could be the consequence of grain coagulations. Recently,
\citet{Ysard18} analyzed the effect of grain growth by comparing the
dust properties derived for isolated grains and aggregates, using Mie
theory and Discrete Dipole Approximation (DDA) calculations. They evidenced a far-infrared (FIR) dust-mass opacity
increase due to the porosity induced by coagulation. Moreover, in the recent Planck Catalogue of Galactic Cold Clumps
\citep{Planck15CC}, the authors determined a mean spectral index of
1.9 when considering all clumps with a good quality of flux, and 2.1 for clumps with temperatures
below 14 K. These values are 
larger than the mean value of $\sim$1.6 obtained for the diffuse
Galactic emission \citep{Planck13XI}. 
\begin{figure}
\begin{center}
\includegraphics[width=8.5cm]{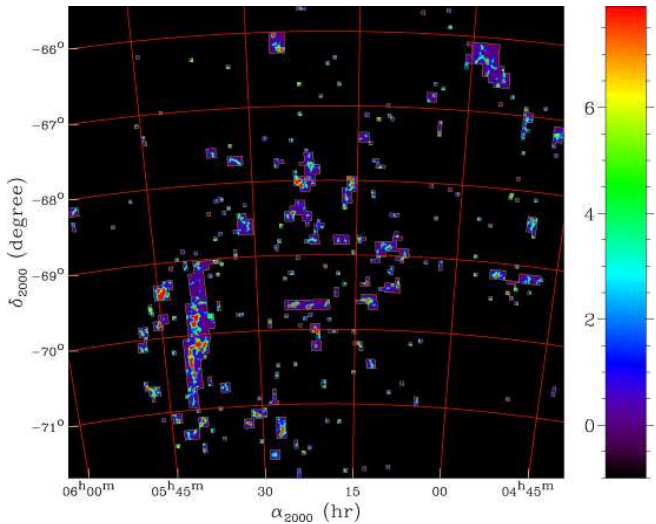}
\caption{Mopra $^{12}\rm CO$ integrated intensity survey (in K km/s) of molecular clouds in the LMC, as part of the MAGMA project \citep{Wong11}. \label{fig_mopra}}
\end{center}
\end{figure}
\begin{figure*}
\begin{center}
\includegraphics[width=17cm]{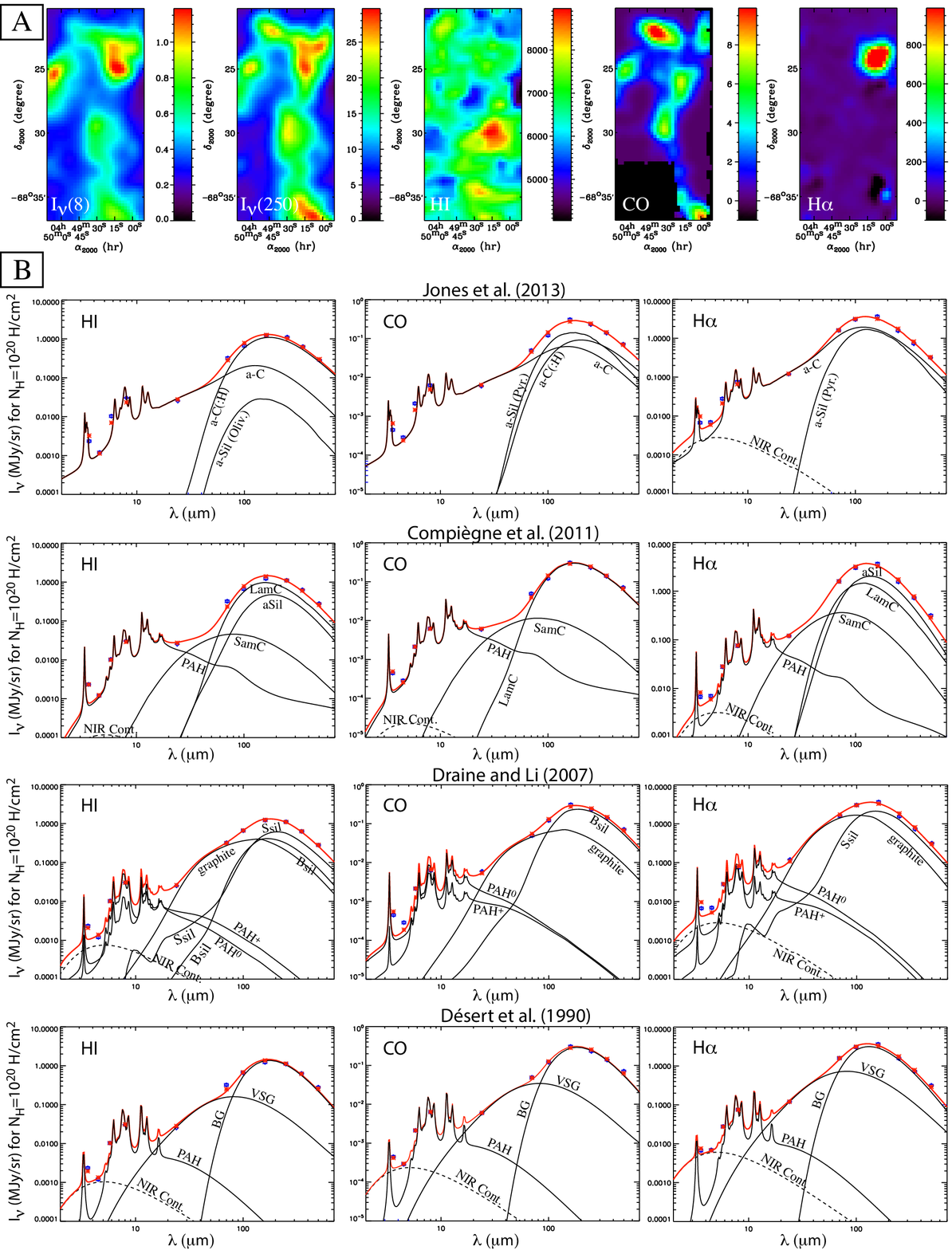}
\caption{Part A: Maps of the dust emission (8$\mic$ and
    250$\mic$ in MJy/sr) and maps
  of the gas components (HI in Jy/beam km s$^{-1}$, CO in K km
  s$^{-1}$, H$\alpha$ in dR) of a molecular cloud of
  our sample \citep[cloud number 5 in][nomenclature]{Fukui08}. Part
  B: SEDs in the various gas phases derived from dust decomposition
  (see Eq. 3) for cloud number 5, fitted with the different dust
  models. Blue diamonds correspond to the data with their 1-$\sigma$ uncertainties, whereas the red
  asterisks and the red continuous lines correspond to the color-corrected brightnesses and the total emission derived from the
  models. \label{fig_sed_4}}
\end{center}
\end{figure*}

\citet{Paradis11a} performed a first statistical analysis of the
272 molecular clouds of the Large Magellanic Cloud (LMC) by
decomposing the dust emission observed with Spitzer
and IRAS into the atomic and molecular gas phases (using ATCA/Parkes
HI and NANTEN $^{12}$CO data). This study did not allow the identification of any
statistical changes in the dust properties between the gas phases, as
opposed to some studies of Galactic molecular clouds \citep{Stepnik03,
  Paradis09, Ysard13}. However, this
analysis was limited in longer wavelengths to 160 $\mic$ and was done at 4$^{\prime}$
angular resolution, corresponding to spatial scales of 60 pc at the
distance of the Large Magellanic
Cloud \citep[$\simeq$50 kpc,][]{Keller06,Feast99}.

In this present paper, we perform a similar analysis as in
\citet{Paradis11a} but at 1$^{\prime}$ angular resolution
(corresponding to spatial scales of $\simeq$ 15 pc), thanks
to the Mopra $^{12}$CO data obtained as part of the MAGMA project \citep{Wong11}. At this
angular resolution we expect to investigate dust properties in cold-enough
environments. Spitzer
data combined to Herschel data allow the spectral 
range to be extended to the submillimeter(submm) domain, crucial to observe any potential changes in
the slope of the dust emission spectra. Moreover, the ionized gas
phase is included in the dust emission decomposition. The spectral
energy distributions (SEDs) are modeled using different dust models:
\citet[][hereafter AJ13]{Jones13}, \citet[][hereafter MC11]{Compiegne11}, \citet[][hereafter DL07]{Draine07}
and an improved version of the \citet[][hereafter DBP90]{Desert90} model. This was done to avoid results that would be strongly model dependent. This work also allows the different dust
models developed to reproduce the diffuse interstellar medium (ISM) to be tested in
several environments characterized by different densities and
temperatures.  
\begin{figure*}
\begin{center}
\includegraphics[width=17cm]{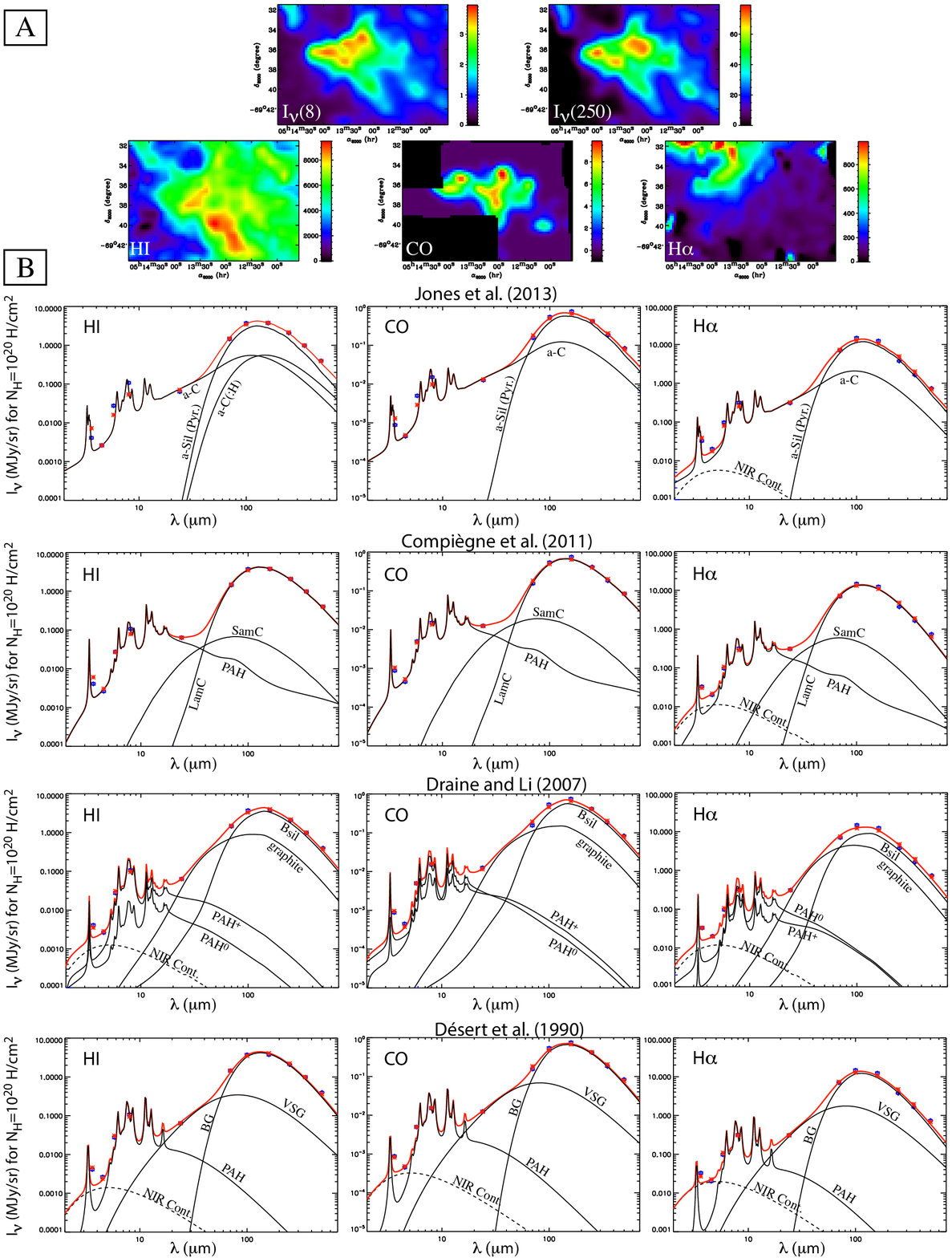}
\caption{Part A: Maps of the dust emission (8$\mic$ and
    250$\mic$ in MJy/sr) and maps
  of the gas components (HI in Jy/beam km s$^{-1}$, CO in K km
  s$^{-1}$, H$\alpha$ in dR) of a molecular cloud of
  our sample \citep[cloud number 71 in][nomenclature]{Fukui08}. Part
  B: SEDs in the various gas phases derived from dust decomposition
  (see Eq. 3) for cloud number 71, fitted with the different dust
  models. Blue diamonds correspond to the data with their 1-$\sigma$ uncertainties, whereas the red
  asterisks and the red continuous lines correspond to the color-corrected brightnesses and the total emission derived from the
  models. \label{fig_sed_52}}
\end{center}
\end{figure*}
After a brief description of the data sets in Section \ref{sec_obs},
we explain in Section \ref{sec_decomposition} the methodology
we adopt to decompose the dust emission in the different gas
phases. Sections \ref{sec_models} and \ref{sec_residus} present the
dust models we use here, as well as their fitting residuals compared
to the observations. In Section \ref{sec_flattening} we analyze the
submm behavior of the SEDs related to the gas phases. We
then describe and discuss the results of this work in Sections
\ref{sec_dustprop} and \ref{sec_discussion}. Conclusions are given in Section \ref{sec_cl}. 
\begin{figure*}
\begin{center}
\includegraphics[width=16cm]{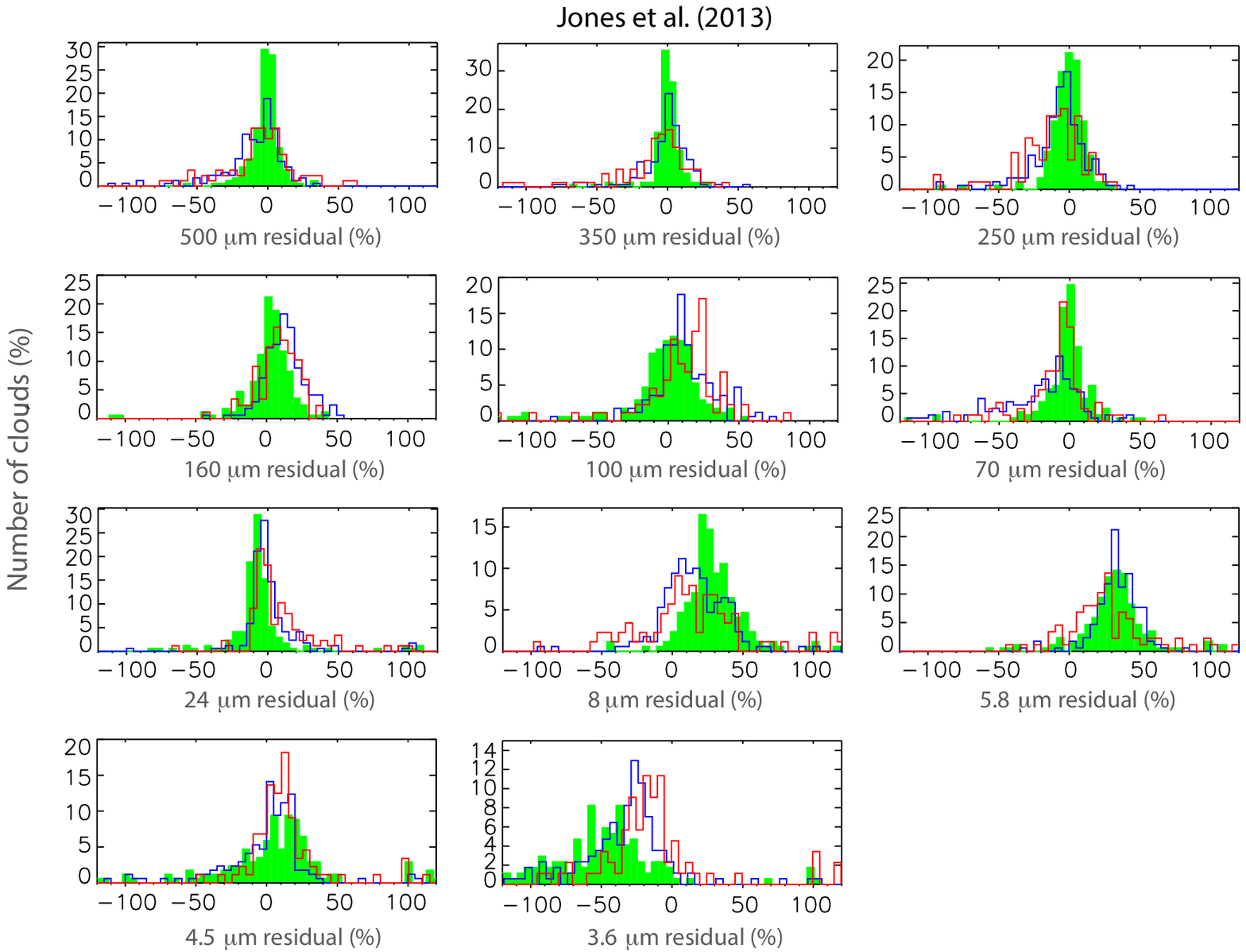}
\caption{Histograms of residuals
  $(I_{\nu}^{\rm obs}-I_{\nu}^{\rm model})/I_{\nu}^{\rm obs}$ derived from AJ13, in the
  atomic (green), molecular (blue), and ionized (red) phases. \label{fig_hist_res_AJ13}}
\end{center}
\end{figure*}
\section{Observations}
\label{sec_obs}
\subsection{Infrared data}
\subsubsection{Spitzer}
In order to trace the IR emission from 3.6 to 160 $\mic$, we use data
from two instruments on board the Spitzer satellite: the Infrared Array
Camera \citep[IRAC;][]{Fazio04} and  the Multiband Imaging Photometer
for Spitzer \citep[MIPS;][]{Rieke04}. These data were obtained as part
of the SAGE (Surveying the Agents of Galaxy Evolution) Spitzer legacy survey \citep{Meixner06}, covering the entire
LMC. The IRAC observed  at  3.6,  4.5,  5.8,  and  8 $\mic$  with  an  angular
resolution  ranging  from  1.6$^{ }$ to 1.9$^{\prime \prime}$.  The MIPS  provided  images at
24,  70,  and  160 $\mic$ at  an  angular  resolution  of  6$^{ }$, 18$^{ }$,
and 40$^{\prime \prime}$, respectively. Only the 24 and 160 $\mic$ MIPS data have been
used whereas MIPS data at 70 $\mic$ have been replaced by the PACS 70
$\mic$ data (see following section). 

We apply the photometric correction to the IRAC maps by multiplying
them by 0.737, 0.772, 0.937, and 0.944 (from 3.6 to 8 $\mic$) to account for the difference between
calibration on point-source and extended sources \citep{Reach05}.
We assume a 10$\%$  calibration uncertainty for all Spitzer data, taking
the repeatability of measurements into account in the
uncertainty. We
then make the quadratic sum of the absolute calibration error and the
statistical errors computed as the standard deviation in a region with low
emission, that is,  the entire region located outside a ring of
4$\degr$ in radius from the center of the maps
 ($\alpha_{2000}=05^h18^m48.0^s$, $\delta_{2000}=-68.7\degr$).

\subsubsection{Herschel}
Dust emission in the FIR to the submm domain was mapped
using the Herschel PACS (70 and 100 $\mic$) and SPIRE (250, 350 and
500 $\mic$) instruments, as part of the Heritage
program \citep{Meixner10}. The angular resolution of the original data
is between 5$^{ }$ and 36$^{\prime \prime}$. We take
absolute uncertainties of 10$\%$ for PACS \citep{Poglitsch10} and
7$\%$ for SPIRE (observer manual v2.4). The white noise computed in
the same way as for Spitzer data is added in quadrature. 
\begin{figure*}
\begin{center}
\includegraphics[width=16cm]{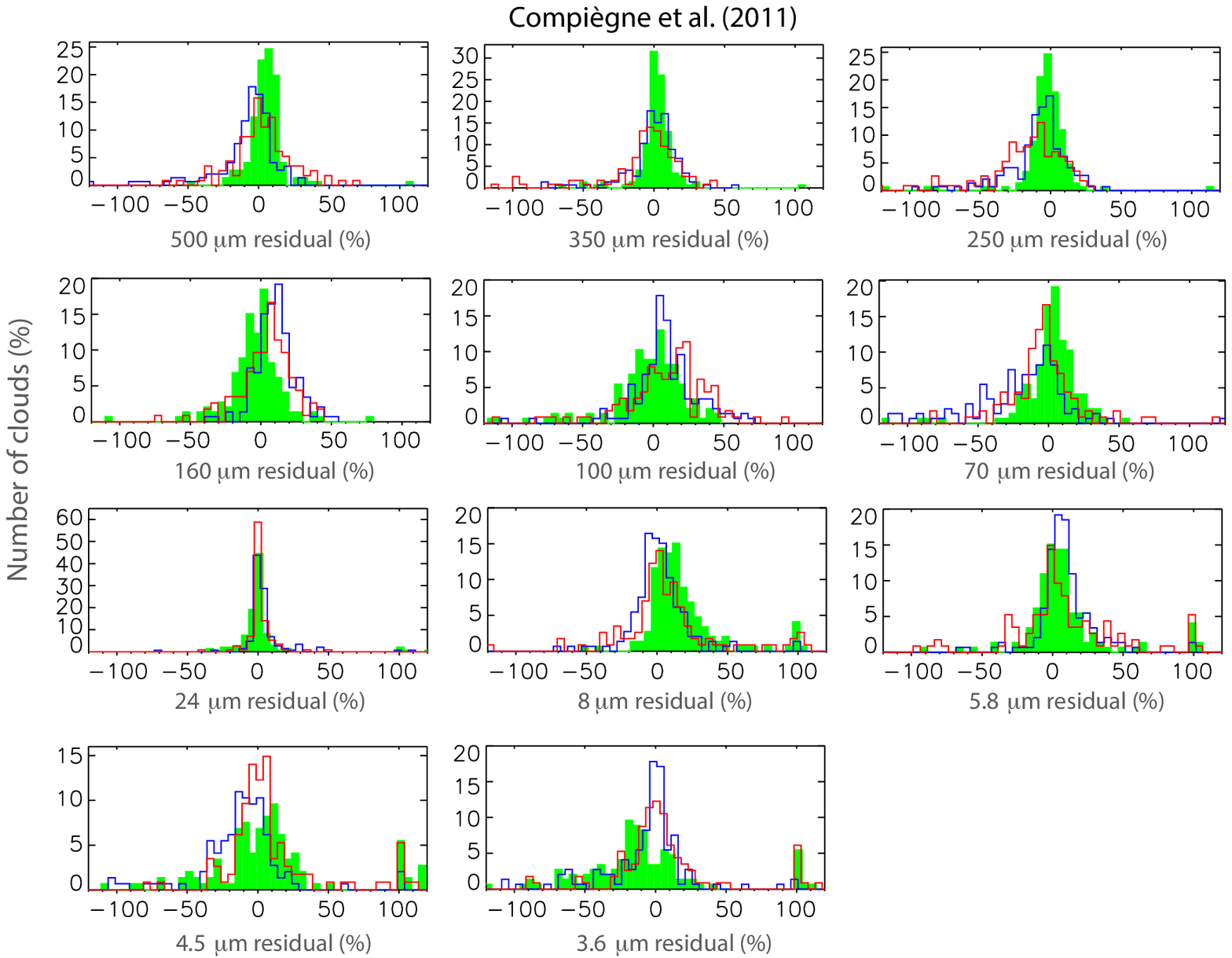}
\caption{Histograms of residuals $(I_{\nu}^{\rm obs}-I_{\nu}^{\rm
    model})/I_{\nu}^{\rm obs}$ derived from MC11, in the different gas phases (same colors as in Fig. \ref{fig_hist_res_AJ13}). \label{fig_hist_res_MC10}}
\end{center}
\end{figure*}
\newline
The OI 63 $\mic$ (or OIII 58 and 88 $\mic$) line could contaminate the 70 and 100 $\mic$
  bands. Its contribution to the MIPS 70 $\mic$ band has been studied
  in the LMC by \citet{Bernard08}. The line should be very bright to
  significantly contribute to the brightness at 70 $\mic$. Some [CII] 158 $\mic$ and [OI] 145
  $\mic$ line emissions should also been visible in the MIPS 160
  $\mic$, which were not observed. We therefore do not expect
  significant contributions from O or C lines.
\begin{table*}
  \caption{Central value ($A_1$) and standard deviation ($A_2$) of Gaussian fits
    ($f(x)=A_oe^{-\left ( \frac{x-A_1}{A_2} \right )^2/2}$) performed on the
  histograms of the model residuals $(I_{\nu}^{\rm
    obs}-I_{\nu}^{\rm model})/I_{\nu}^{\rm obs}$ in percent (see
    Figs. 4 to 7) for each phase of
  the gas. \label{table_excess}}
\begin{center}
\begin{tabular}{lcccccccccccc}
\hline
\hline
 \multicolumn{12}{c }{ \citet{Jones13} } \\
 & & 3.6 $\mic$ & 4.5 $\mic$ & 5.8 $\mic$ & 8 $\mic$& 24 $\mic$& 70
 $\mic$& 100 $\mic$ & 160 $\mic$ & 250 $\mic$ & 350 $\mic$ &500 $\mic$ \\
\hline
HI & A$_1$  & -47.74& 10.29 &31.49&25.21 &-7.18&-2.76&1.43&2.72&-0.96&0.56&0.01\\
 & A$_2$ & 26.58 & 18.56&12.39&13.73& 5.27&6.47&14.59&9.01&8.88&5.16&5.49\\
\hline 
CO &A$_1$ &-29.88 & 5.93&35.86 &13.96 &-5.61&-13.67 &8.66&11.12& -5.52&1.53&-5.15\\
& A$_2$ & 15.56 & 11.92 &11.62&19.22&5.79&17.61&13.02&11.78&11.01&8.95& 12.53\\
\hline
H$\alpha$&A$_1$ &-19.56 & 6.94&21.24&10.38&-3.67 &-6.57 &14.83 & 12.53&-10.16&-2.19& -3.98 \\
& A$_2$ & 14.20 &10.61 &17.54 &23.39& 8.45 &7.70 &16.83&12.90&17.59&11.85&12.28\\
\hline
 \multicolumn{12}{c }{ \citet{Compiegne11} } \\
\hline
HI & A$_1$  & -14.74& 3.55&0.93&8.10&-1.56&5.56&2.40&-3.48&-3.52&1.33&3.68 \\
 & A$_2$ & 19.55 &17.00 &10.36&11.59&3.25&10.45&16.82&11.54&7.79&5.65& 6.99\\
\hline 
CO&A$_1$ & -1.62& -9.34&7.01&-1.22&-1.72&-12.04&9.28&7.35&-6.23&1.27& -4.00\\
& A$_2$ & 7.92& 14.95&8.59&11.07&3.32&18.87&12.04&10.66&10.99&10.55&9.43 \\
\hline
H$\alpha$&A$_1$ &-1.98 & -2.03&-0.58&0.80&-1.55&-5.62&11.65&5.56&-11.17&1.77& -0.144\\
& A$_2$ & 12.82& 10.50&13.26&11.79&2.67&11.38&21.88&13.73&19.34&12.22& 13.90\\
\hline
 \multicolumn{12}{c }{ \citet{Draine07} } \\
\hline
 HI & A$_1$  & 5.41&-1.94 &-2.08&0.74&-1.85&0.27&9.07&-2.14&-5.96&-2.05& 4.33\\
 & A$_2$ & 24.05& 19.16&9.84&7.43 &5.79&7.60&13.27&10.93&7.73&5.42&8.08 \\
\hline 
CO&A$_1$ & -16.29 &21.83&3.69&-6.81&9.01&-22.00&12.10&9.58&-5.77& 2.32&-5.34\\
& A$_2$ & 9.86& 18.71&7.82&7.37&10.34&18.47&11.32&11.33&6.24&9.35&13.13\\
\hline
H$\alpha$&A$_1$ &-15.27 &10.02&0.98&-3.55&5.25&-10.38&18.32&8.09 & -8.23&-0.15&4.55\\
& A$_2$ & 20.87&18.89&13.70&10.78&8.19&12.20&15.27&12.29&16.54&12.91&10.70\\
\hline
 \multicolumn{12}{c }{ \citet{Desert90} } \\
\hline
HI & A$_1$  &4.87 & -3.72&-3.45&0.14&-2.72&6.26&3.87&-5.62&-7.22&0.29&10.38\\
 & A$_2$ & 13.15&12.27 &8.41&7.98&4.25&13.07&15.07&13.24&8.71&5.87& 9.56\\
\hline 
CO &A$_1$ & 12.45& -12.84&4.10&-7.66&0.83&-14.99&10.05&8.71&-7.12&1.96&-0.45 \\
& A$_2$ & 9.02& 7.53&7.42&7.19&4.86&17.35&11.46&12.31&12.66&10.75& 14.99\\
\hline
H$\alpha$&A$_1$ &8.96 &-5.68 &1.30&1.76&-2.55&-5.85&9.41&1.81&-8.01&3.21& 9.55\\
& A$_2$ & 12.06 &8.65&12.13&10.20&3.67&8.63&17.46&15.88&16.88&14.30&18.92\\
\hline
\hline
\end{tabular}
\end{center}
\end{table*}

\subsection{Gas tracers}
\subsubsection{HI emission}
As a tracer of the atomic gas, we use the \citet{Kim03} 21 cm map, at a spatial
resolution of 1$^{\prime}$, which is a combination
of interferometric data from the Australia Telescope Compact Array
(ATCA; 1$^{\prime}$), and the Parkes antenna \citep[15.3$^{\prime}$;][]{Staveleysmith03}. The HI data were integrated in the velocity
range 190 km s$^{-1}$$<\rm V_{LSR}<386$ km s$^{-1}$.
The HI integrated intensity map has been converted to HI column
density ($N\rm_H^{HI}$) by
applying the standard conversion factor, $X\rm_{HI}$, equal to $1.82\times 10^{18}$
$\rm H/cm^2/(K\,km\,s^{-1}),$ \citep{Spitzer78, Lee15}.
\begin{figure*}
\begin{center}
\includegraphics[width=16cm]{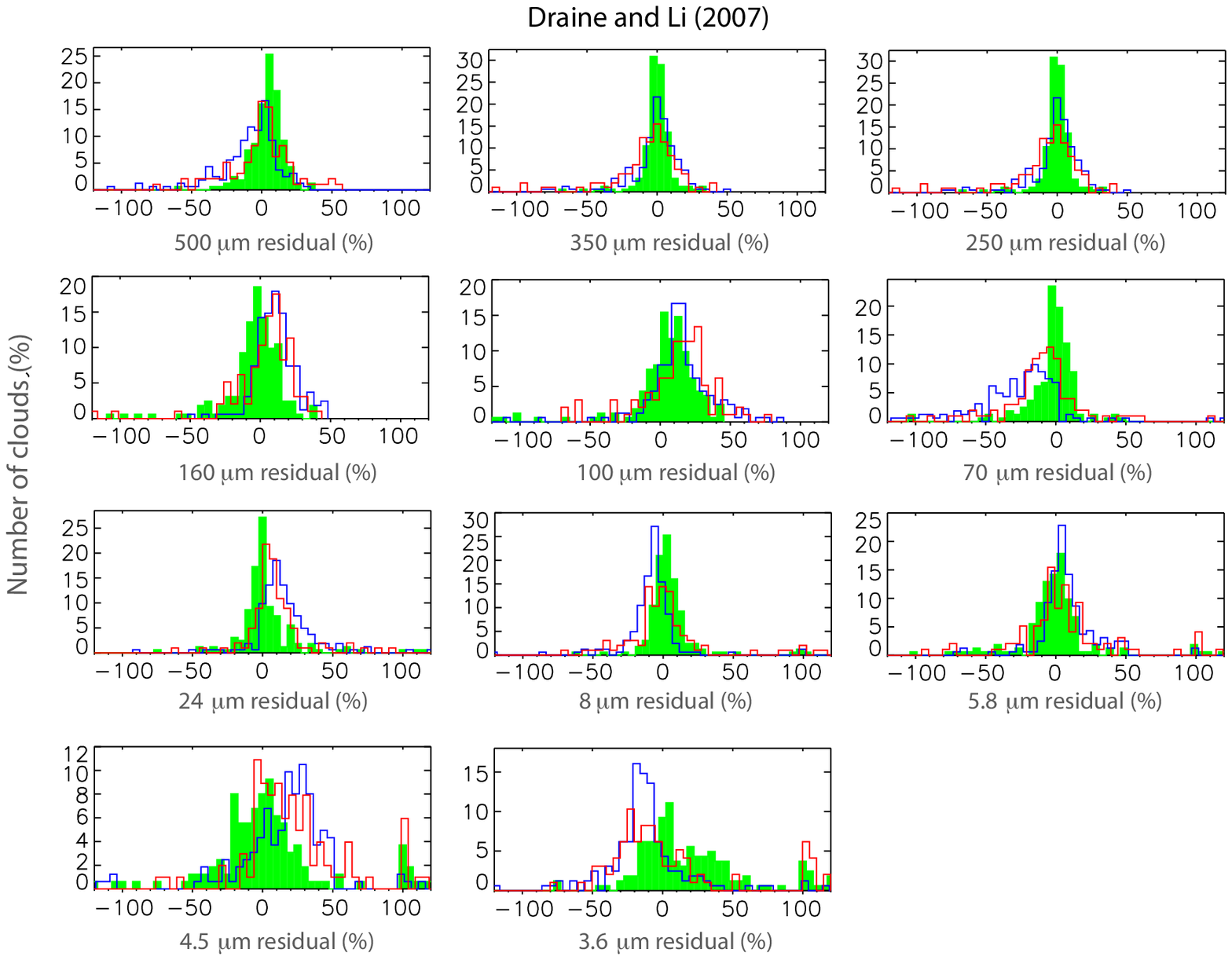}
\caption{Histograms of residuals $(I_{\nu}^{\rm obs}-I_{\nu}^{\rm model})/I_{\nu}^{\rm obs}$ derived from 
  DL07, in the different gas phases (same colors as in Fig. \ref{fig_hist_res_AJ13}). \label{fig_hist_res_DL07}}
\end{center}
\end{figure*}
\begin{figure*}
\begin{center}
\includegraphics[width=16cm]{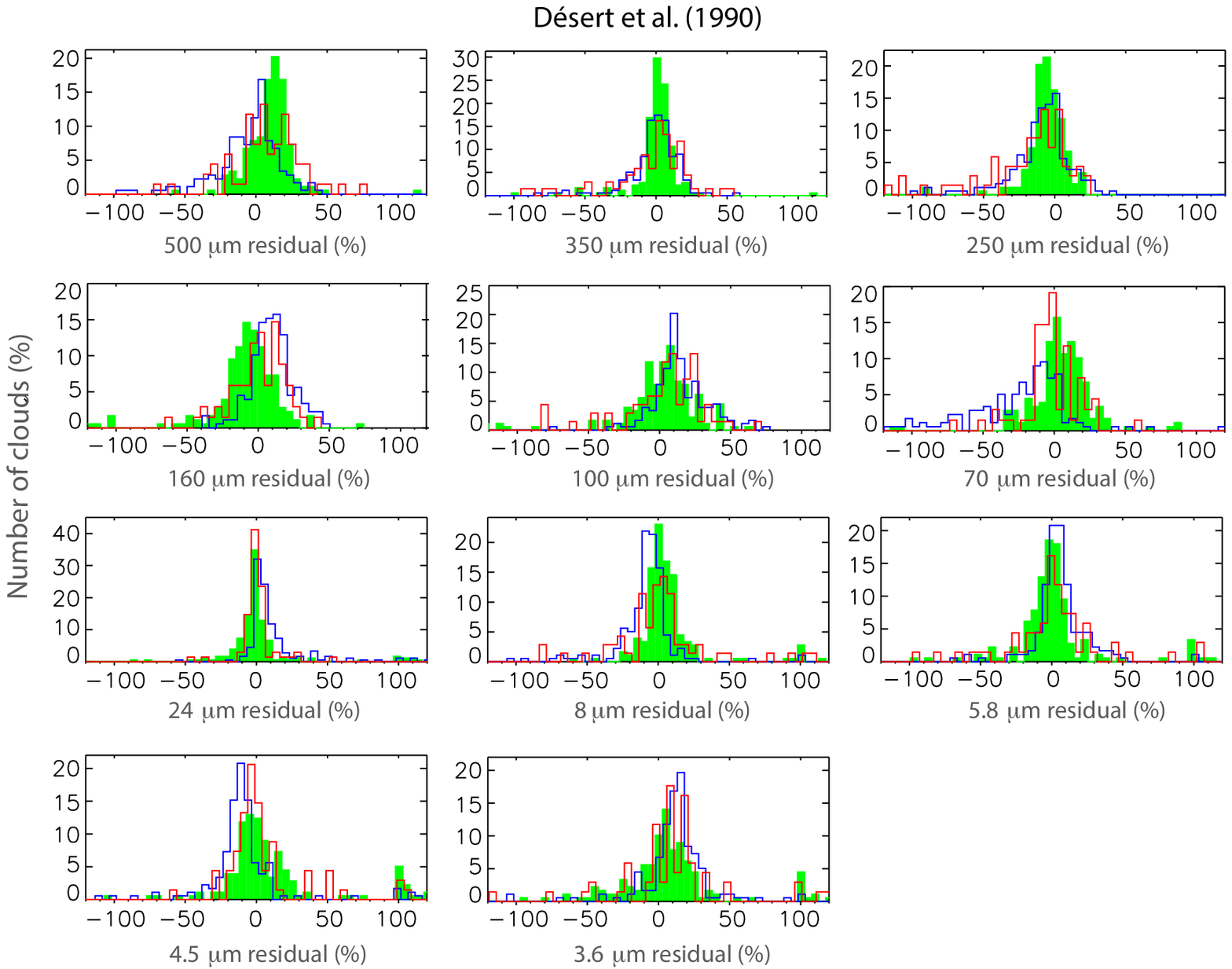}
\caption{Histograms of residuals $(I_{\nu}^{\rm obs}-I_{\nu}^{\rm model})/I_{\nu}^{\rm obs}$ derived from 
  DBP90, in the different gas phases (same colors as in Fig. \ref{fig_hist_res_AJ13}). \label{fig_hist_res_DBP90}}
\end{center}
\end{figure*}

\subsubsection{CO emission}
As a tracer of the molecular gas, we use the latest release of the data
obtained with the 22 m Mopra
telescope of the Australia Telescope National Facility, as part of
the MAGMA project \citep{Wong11}. The MAGMA LMC survey is a follow-up
to the NANTEN survey at a resolution of $\simeq$1$^{\prime}$ (see Fig.
\ref{fig_mopra}). The integrated intensity maps ($W\rm_{CO}$) are converted to molecular
column densities using the relation:
\begin{equation}
N_{\rm H_2}=X_{\rm CO}W_{\rm CO}
,\end{equation}
with $X\rm_{CO}$ being the CO-to-H$_2$ conversion factor. \citet{Hughes10}
found an average value of 4.7 $\times 10^{20}$ $\rm
H/cm^2/(K\,Km\,s^{-1})$ for the LMC MAGMA clouds. \citet{Leroy11}
determined a value of 3 $\times 10^{20}$ $\rm
H/cm^2/(K\,km\,s^{-1})$ in the LMC, whereas \citet{Roman-Duval14} found
upper limits to be 6 $\times 10^{20}$ $\rm
H/cm^2/(K\,km\,s^{-1})$. Taking the dispersion of values into account,
we decided to adopt an intermediate value equal to 4 $\times 10^{20}$ $\rm
H/cm^2/(K\,km\,s^{-1})$.
\begin{figure}
\begin{center}
\includegraphics[width=8cm]{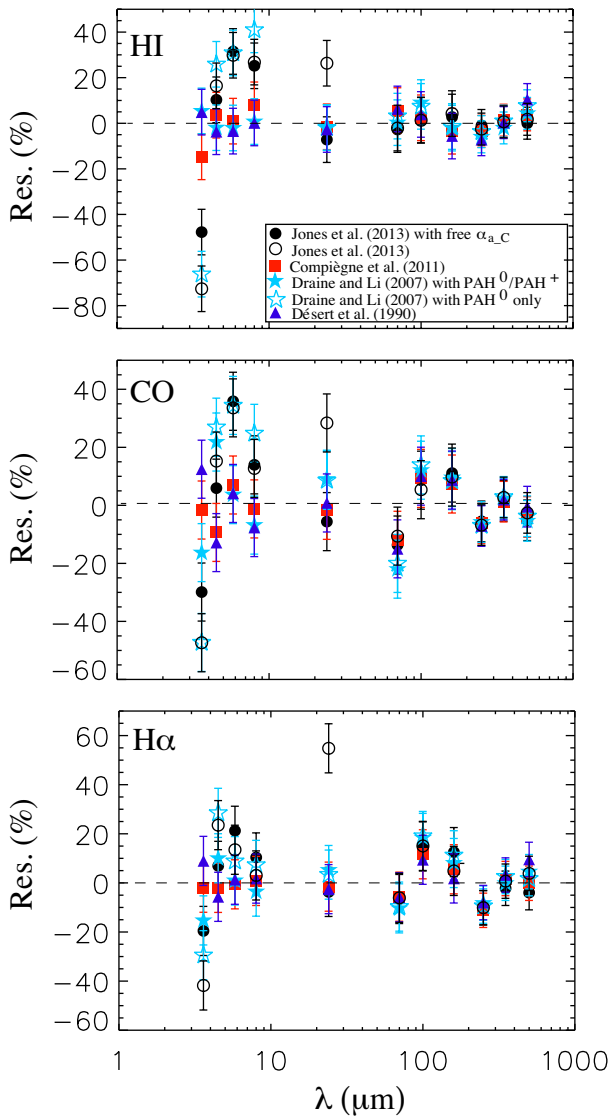}
\caption{Average residuals $A_1$ based on Gaussian fits ($f(x)=A_oe^{-\left ( \frac{x-A_1}{A_2} \right )^2/2}$) presented
  in Figs. \ref{fig_hist_res_AJ13} to \ref{fig_hist_res_DBP90} and
  summarized in Table \ref{table_excess}, as a function
  of wavelength. Error bars
  correspond to the data calibration uncertainties (10$\%$ for
  Spitzer and Herschel/PACS data, and 7$\%$ for Herschel/SPIRE data). The
  dashed line shows the absence of residual (0$\%$). \label{fig_res}}
\end{center}
\end{figure}

\begin{table}
\caption{Central value ($A_1$) and standard deviation ($A_2$) of Gaussian fits
  ($f(x)=A_oe^{-\left ( \frac{x-A_1}{A_2} \right )^2/2}$)  performed on
  the histograms of $X_{ISRF}$ for the atomic and molecular phases for
  each dust model. \label{table_xisrf}}
\begin{center}
\begin{tabular}{lccccc}
\hline
\hline
  & & AJ13 & MC11 & DL07 & DBP90 \\
\hline
HI       & A$_1$ &3.107 &1.313& 1.004& 1.925 \\
       & A$_2$ & 1.659 & 0.861&0.669 & 1.268 \\
\hline
 CO      & A$_1$ &1.260 &1.055&0.441&1.034 \\
       & A$_2$ &1.062 &1.196&0.945&1.094 \\
  \hline
  \hline
  \end{tabular}
\end{center}
\end{table}
\subsubsection{H${\alpha}$ emission}
The warm ionized gas, with an electron temperature $\rm T_e \simeq
10^4$ K, emits both recombination lines and radio free-free
continuum. The H$\alpha$ line is the brightest of the recombination lines. To trace the ionized gas we use the
Southern H-Alpha Sky Survey Atlas \citep[SHASSA,][]{Gaustad01} data set which
covers the southern hemisphere ($\delta <15\degr$). The angular
resolution of the data is $0.8^{\prime}$, and the sensitivity is 2 Rayleigh per pixel (1
Rayleigh=10$^6$ photons cm$^{-2}$ s$^{-1}$). By assuming a constant electron
density $n_e$ along each line of sight, we derive the $\rm H^+$
column density with the relation \citep{Lagache99}: 
\begin{equation}
\label{eq_halpha}
\frac{N{\rm_H^{H^+}}}{{\rm H\,cm^{-2}}}=1.37 \times 10^{18}
\frac{I_{\rm {H
    \alpha}}}{\rm R} \frac{n\rm_e}{\rm cm^{-3}}
,\end{equation}
where 1 Rayleigh=2.25 pc cm$^{-6}$ for T$\rm _e$=8000 K
\citep[see, for instance][]{Dickinson03}. 

To determine the electron density we first compute the mean emission
measure (EM; $1EM=2.75T_4^{0.9}I(H\alpha)$ cm$^{-6}$ pc) for all
pixels associated with molecular clouds, by assuming an electron
temperature of 8000 K. As in \citet{Paradis11b} we
adopt a vertical extent of H$^+$ of 500 pc. We then obtain an electron
density $n\rm_e$ close to 1 cm$^{-3}$. This value is likely an upper
limit since the vertical extent of H$^+$ is approximate and could be as
high as the estimated scale height in our Galaxy ranging between 1000 and 1800 pc.
By comparison, \citet{Paradis11b} derived a value of $n\rm_e$=0.055 cm$^{-3}$
for the diffuse ionized gas of the LMC, $n\rm_e$=1.52 cm$^{-3}$ for
typical HII regions and $n\rm_e$=3.98 cm$^{-3}$ for bright HII
regions. 

\subsection{Additional data processing}
\subsubsection{Galactic foreground}
Infrared data of the LMC can be significantly affected by
Galactic foreground emission. The IR foreground contribution at all
Spitzer and Herschel
wavelengths was subtracted from the maps using the HI
foreground map constructed by
\citet{Staveleysmith03} at an angular resolution of 14$^{\prime}$. 
Parkes HI data have been integrated in the velocity range
$-64<v_{LSR}<100$ Km s$^{-1}$, excluding emission coming from the
LMC itself. The IR foreground contribution from 3.6 to 100 $\mic$ has
been taken from \citet{Bernard08}, using the IR brightness values corresponding to the best fit of
the average high-latitude Galactic emission SED of Dwek et al.
(1997). The contribution in the Herschel SPIRE bands has been derived
from a modified black body.   

\subsubsection{Star removal}
 IRAC maps are highly affected by stellar emission. However, to perform
a rigorous analysis, we removed the stellar contribution in all
Spitzer and Herschel maps. We followed the exact same procedure as
described in \citet{Bernard08}. 

\subsubsection{Convolution to 1 arcminute}
To compare different data sets, all maps need to be convolved to a common resolution, set to the lowest resolution
of the data sets, which is 1${^{\prime}}$ here. We used a Gaussian
kernel with $\theta_{\rm FWHM}$ $= \sqrt{ \left(
    (1^{\prime})^2-(\theta{_{\rm FWHM}^{\rm d}})^2 \right) }$ 
, with $\theta{\rm_{FWHM}^{\rm d}}$ the original
resolution of the maps.  

\section{Decomposition of the dust emission}
\label{sec_decomposition}
We perform correlations between IR emission and gas tracers for
each molecular cloud. We use the same cloud classification as
described in the \citet{Fukui08} catalog, which accounts for 272
molecular clouds. Each cloud observed with Mopra is extended by 4
pixels (with a pixel size of 14$^{\prime \prime}$) to include HI and H$\alpha$ gas surrounding the clouds. 
We then decompose the IR--submm emission at each
wavelength (from 3.6 $\mic$ to 500 $\mic$), such as:
\begin{equation}
I_{\nu}(\lambda)=aN{\rm _H^{HI}}+bN{\rm _H^{CO}}+cN{\rm _H^{H^+}}+d
,\end{equation} 
where the $a$, $b$, $c$, parameters denote the emissivity associated with each phase
of the gas, and $d$ is a constant that could represent an additional gas
phase such as the dark gas component, and/or possible offsets in the
data. We obtain a SED representative of
each phase of the gas, for each molecular cloud.
\begin{figure*}
\begin{center}
\includegraphics[width=14cm]{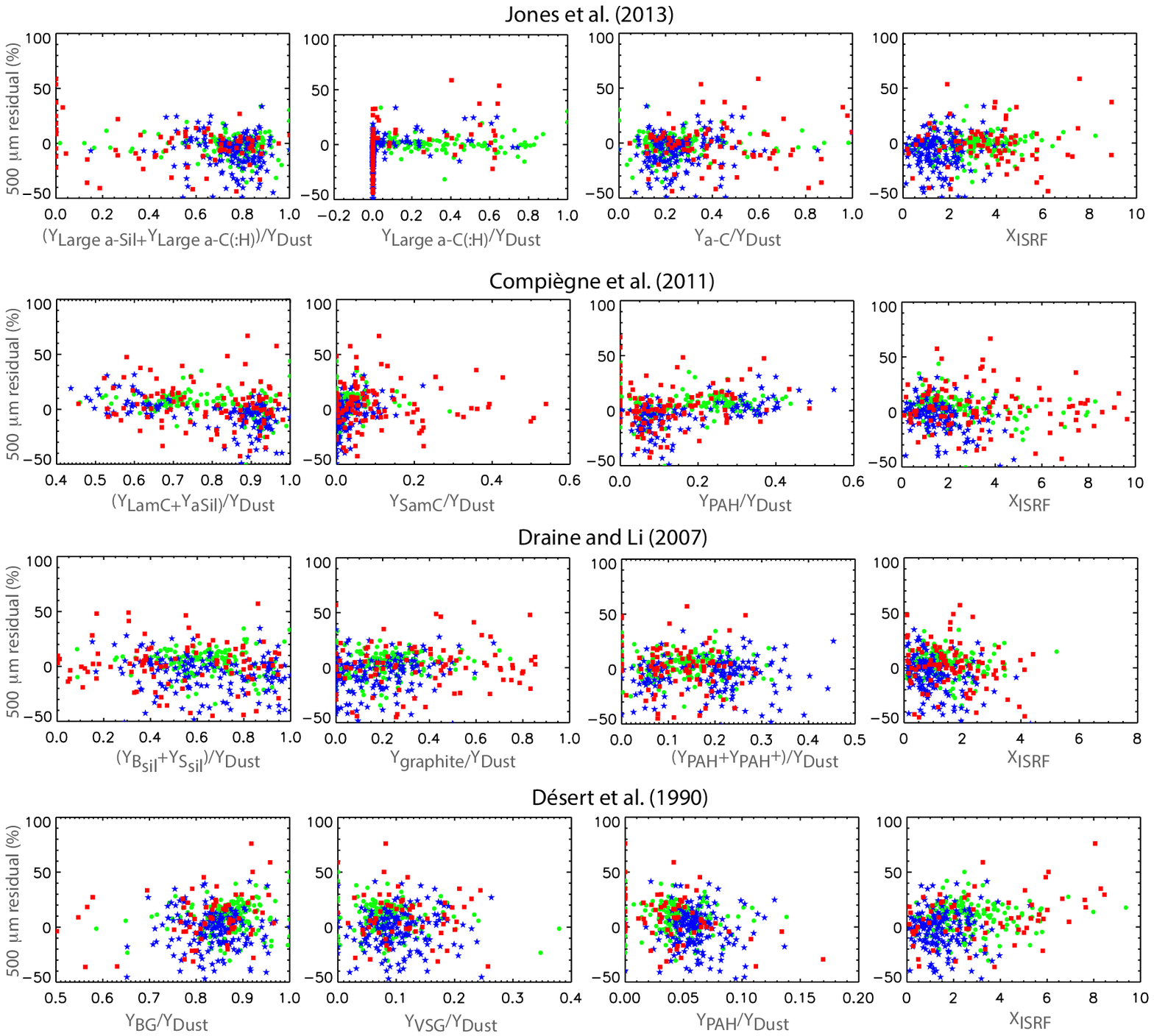}
\caption{500 $\mic$ residual as a function of the model parameters (relative dust abundances and intensity of the radiation field;
    see Sect. 4.2) for
  the different gas phases (atomic in green, molecular in blue, and
  ionized in red).  \label{fig_excess_ab}}
\end{center}
\end{figure*}

In \citet{Paradis11a} the dark gas component had been included in the
decomposition. We reconstructed a dark gas map at a resolution of the 1$^{\prime}$ following the method of \citet{Bernard08} and adopting the
$X_{\rm CO}$ value defined in our analysis, and
performed the decomposition. This new decomposition significantly reduced the
quality of the obtained SEDs, that is, it gives more negative correlations. 
It significantly reduces the statistics of
this analysis by excluding many clouds. Moreover, the dark gas component is highly correlated to the
atomic gas since this component is used when constructing the dark gas
map. For all these reasons, we decided to present results obtained when
we do not include the dark gas. 
The main impact of the absence of dark
gas in the decomposition is on the absolute levels of gas column densities
(especially in the atomic phase), and as
a consequence in the absolute levels of the
dust abundances. The effect does not have significant consequences for the dust relative
abundances. 
We carefully checked the consistency of all our results with and
without the dark phase and can ensure that the conclusions are
identical (see Sect. \ref{sec_ab} and the Appendix for a comparison of the results). 

\begin{figure*}
  \begin{center}
\includegraphics[width=16cm]{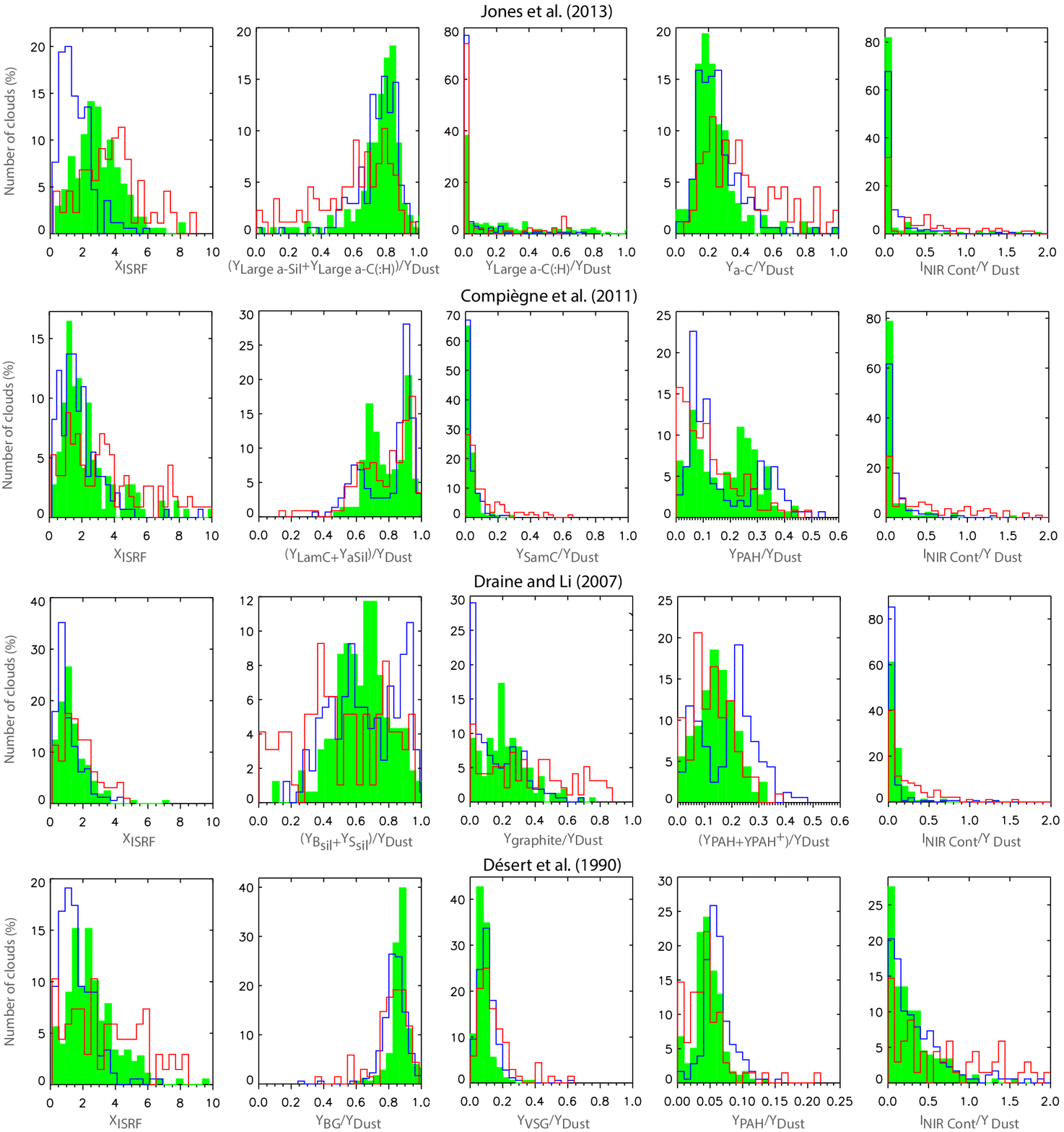}
\caption{Histograms of the parameters derived for each dust model 
    (relative dust abundances, intensity of the radiation field, and relative intensity of the NIR
    continuum; see Sect. 4.2) in
  the different gas phases (atomic in green, molecular in blue, and
  ionized in red). We notice a decrease of the intensity of the radiation field in
the molecular phase, as well as an increase in the ionized phase, as
compared to the atomic phase. We
also observe an increase of the PAH relative abundance in the
molecular phase (through a positive shift of the
entire or part of the histogram in the molecular phase as compared to
the atomic phase), and an increase of the VSG relative abundance in the
ionized phase. The distribution of the NIR-continuum is more
extended in the ionized phase than the ones in the atomic and molecular phases.\label{fig_param_AJ13}}
\end{center}
\end{figure*}

\section{Modeling}
\label{sec_models}
\subsection{Description of the models}
The DustEM package\footnote{see
  https://www.ias.u-psud.fr/DUSTEM/ for the fortran code and
  http://dustemwrap.irap.omp.eu/ for the IDL wrapper} offers the possibility to use four
different dust models:\\
- A version of the \citet{Desert90} model, that consists in
three dust components: the polycyclic aromatic hydrocarbons (PAHs), the
very small grains (VSGs) probably carbon dominated, and the big grains
(BGs) probably made of a size distribution of silicates with some dark 
refractory mantles. The  original  model was  updated to reproduce  the  actual  shape  of  the  PAH
emission features as derived from ISO and Spitzer spectroscopy
measurements  and  to  include  the  17$\mic$  feature  attributed to
PAHs.   \\
- The \citet{Draine07} model, also called silicate-graphite-PAH model, that assumes a mixture of carbonaceous
and amorphous silicates grains, including different amounts of
PAH material. \\
- The \citet{Compiegne11} model that comprises PAHs, small ($\it a$$<$10 nm)
and large ($\it a$$>$10 nm) amorphous
carbons (SamC and LamC), and large amorphous silicates (aSil); \\
- The THEMIS model \citep{Jones13}, that provides an evolutionary scenario for the dust evolution in the ISM. This model comprises two dust
components: a population of carbonaceous grains of
amorphous and aliphatic nature (a-C(:H)) for the
large grains ($\it a$$>$20 nm) and of more aromatic nature (a-C) for the smallest ones ($\it a$$<$20nm), and a population of large amorphous silicate grains containing nanometer scale inclusions
of FeS (large a-Sil). These two grain populations are covered by an aromatic
mantle in the diffuse ISM and in photodissociation regions (PDRs). In dense regions, an additional mantle of more or less aliphatic matter is
accreted on top of the aromatic mantle. 

Depending on the model, the mid-infrared (MIR) emission (20-60 $\mic$) is mainly
dominated by VSGs (DBP90), graphites (DL07), and SamC (MC11). With AJ13,
the amorphous carbons ($a-C$), that also include PAHs, has a continuous
dust-size distribution that reproduces the MIR emission, and can in some cases dominate the FIR emission if adopting a different
power law in the size distribution as compared to the one used to
reproduce the Galactic diffuse ISM.
 
The FIR--submm emission is described by BGs (DBP90), amorphous
silicates (DL07), large amorphous carbons, and large
silicates (MC11). With AJ13, by adopting the same
dust size distribution as in our Galaxy, the large a-C(:H) and
silicate grains dominate the emission at long wavelengths.
For simplicity, we adopt a single name for the different dust
components of the models: PAHs, VSGs (MIR emission), and BGs
(FIR--submm emission).  

\subsection{Fitting}
Each SED is modeled using the DustEM Wrapper package. 
These models were first developed to reproduce the diffuse
ISM of our Galaxy. We  tested each of these models 
(with Galactic parameters) on the LMC clouds by
only constraining the mass abundances relative to hydrogen of the different
dust populations ($Y_i$), the intensity of the NIR continuum ($I_{NIR\,Cont}$), and the
intensity of the radiation field ($X_{ISRF}$). We consider a
  single intensity of the radiation field (RF). This assumption seems
  reasonable since this analysis focuses on molecular clouds which
  are expected to be ``cold'', mainly in the atomic and molecular
  phases. As a consequence, a mixture of RF is not
 required. The ionized gas phase
  surrounding the clouds can belong to three different categories:
  diffuse ionized gas, and typical or bright HII regions. As described
  in \citet{Paradis11b}, the results obtained assuming a composite
 RF substantially corroborate those derived with a
  single RF. However, it appears that a composite
  RF does not always give a better fit than a single
  RF, and in particular does not
reproduce the 70 $\mic$ excess evidenced in the ionized
  phase. For instance, in some cases, modeling with DBP90 including a composite
 RF exhibits an important lack of emission at 70 $\mic$,
  as well as a higher $\chi^2$, as compared to a single RF
  modeling \citep{Paradis11b}. For low temperatures, the use of a composite RF also engenders higher $\chi^2$ \citep{Paradis12}. For all these
reasons, we decided to adopt a single RF. 

All the DustEM models
that include a separate PAH component, that is, DBP90, DL07, and MC11, offer the possibility to have a 
mixture of PAHs depending on their ionization degrees. We have tested
all models with full neutral PAHs and with both neutral and ionized
PAHs. Only DL07 gives significantly better
results with a mixture of PAHs, whereas the presence of ionized PAHs
in the other models does not improve the quality of the
fits. With AJ13, adopting the same dust-size distribution
for the a-C component as in our Galaxy does not give satisfactory
results \citep[see also][]{Chastenet17}, it reproduces neither the near-infrared (NIR) nor the MIR emission. Therefore, the
slope of the power law in the a-C dust size distribution is left as a
free parameter. \\
A NIR-continuum described by a black body at a temperature of 1000 K is added to the
modeling, as in previous studies \citep[see e.g.,][]{Flagey06, Bernard08, Paradis11a}. Its origin is unknown but is required in addition to the
PAH components in most of the cases to reproduce the NIR
data.
To summarize, for each model we leave the following free parameters
in the fits:
\begin{itemize}
\item{DBP90 (5 free parameters): $Y_{\rm PAH}$, $Y_{\rm VSG}$, $Y_{\rm
      BG}$, $X_{\rm ISRF}$ and
$I_{\rm NIR\,Cont}$}
\item{DL07 (7 free parameters): $Y_{\rm PAH^0}$, $Y_{\rm
      PAH^+}$$Y_{\rm graphite}$,
$Y_{\rm B_{sil}}$ for big silicates, $Y_{\rm S_{sil}}$ for small
silicates, $X_{\rm ISRF}$ and
$I_{\rm NIR\,Cont}$}
\item{MC11 (6 free parameters): $Y_{\rm PAH}$, $Y_{\rm SamC}$ for small amorphous
carbons, $Y_{\rm LamC}$ for large amorphous carbons, $Y_{\rm aSil}$ for
amorphous silicates, $X_{\rm ISRF}$ and
$I_{\rm NIR\,Cont}$}
\item{AJ13 (7 free parameters): $Y_{\rm a-C}$ for small aromatic carbons,
$Y_{\rm Large\,a-C(:H)}$ for large aliphatic Carbons,
$Y_{\rm Large\,a-Sil}$ for large amorphous silicates (including two
silicate components: $Y_{\rm Oliv.}$ for olivine-type
and $Y_{\rm Pyr.}$ for pyroxene-type), $\alpha_{\rm a-C}$ for the
slope of the power-law in the a-C dust size distribution, $X_{\rm ISRF}$ and
$I_{\rm NIR\,Cont}$}. 
\end{itemize}
Typical SEDs obtained in the phases of two molecular clouds fitted
with the models are presented in Figs. \ref{fig_sed_4} and \ref{fig_sed_52}. 

The lack of data between 24 and 70 $\mic$ is problematic
  in the sense that there are no further constraints to characterize the VSG component over this wavelength range. The models have also different
  predictions in the submm depending on their emissivity spectral
  index. For instance, both previous figures evidence distinct
  brightness predictions in the submm at 700 $\mic$ for the same phase. Additional
  data in the mm range would be useful to constrain and improve the
  existing models. However, today, including mm data from space instruments in the SED directly
  impacts the angular resolution which has to be degraded.

\subsection{Cloud selection}
\label{sec_selection}
The decomposition of the dust emission does not give satisfactory results in some
cases. For instance, some decompositions can lead to negative
correlations, that can result from low negative emission in some
pixels after
foreground subtraction or star removal,  or from possible correlation between the gas
phases for example. Data at 100 and 160 $\mic$ are both crucial to derive
the dust temperature and to determine the dust abundances. We
therefore only consider SEDs with
positive values at 100 and 160 $\mic$ in both the atomic and
the molecular phases. This criterion induces the rejection of 90 clouds. The same criterion
in the ionized phase leads to the removal of 40 additional clouds. We
therefore consider a total of 182 clouds for the atomic and
molecular phases and 142 clouds for the ionized phase. From these
clouds we also exclude those for which SED fitting results in substantial bad fits. This last selection depends on the model used, and can
remove 20$\%$ of the clouds in the case of MC11, whereas it concerns
10$\%$ of the clouds with DL07, 5$\%$ and 2$\%$ with AJ13 and DBP90.

\section{Modeling residuals}
\label{sec_residus}
Distribution of the modeling residuals compared to observations
(Residual$=(I_{\nu}^{\rm obs}-I_{\nu}^{\rm model})/I_{\nu}^{\rm obs}$) are presented in Figs.
\ref{fig_hist_res_AJ13}, \ref{fig_hist_res_MC10},
\ref{fig_hist_res_DL07}, and \ref{fig_hist_res_DBP90} for each
model. We make a Gaussian fit to the residuals with the
following function:
\begin{equation}
f(x)=A_oe^{-\left ( \frac{x-A_1}{A_2} \right )/2}
.\end{equation}
The $A_0$, $A_1$, and $A_2$ parameters correspond to the amplitude,
the central value, and the standard deviation of the Gaussian. Values of $A_1$ and $A_2$ are summarized in Table \ref{table_excess},
and $A_1$ parameters are plotted in Fig. \ref{fig_res}. In this
latter figure, we also plot results of the fits obtained with 
AJ13 when taking the same size distribution of the
smallest grains as in our Galaxy, and obtained with 
DL07 when considering full neutral PAH component for comparison. DL07 clearly requires the mixture
of neutral and ionized PAHs to improve the fits between 3.6 and 8
$\mic$. The 24 $\mic$ can be reproduced within 15$\%$ by all the models, except with AJ13 adopting the Galactic size distribution of the
a-C component.
We clearly see that the models are more or less in agreement with each
other within 10$\%$ in the
FIR--submm (70 - 500 $\mic$) domain and give reduced residuals compared
to the NIR domain. From Table \ref{table_excess}, we can see that for wavelengths below 24 $\mic$ the dispersion between the
different results obtained with the models is large and can be as high
as 52$\%$ at 3.6 $\mic$, 35$\%$ at 4.5 and 5.8 $\mic$, and 25$\%$ at 8
$\mic$. It also shows the highest residuals at 24 and 70 $\mic$ compared to the
other models. AJ13 does not give satisfactory results in the PAH bands
with residuals reaching values up to 35$\%$ at 5.8$\mic$ and -47$\%$
at 3.6$\mic$ for instance, significantly larger than the calibration uncertainties
(10$\%$). However, this model has the best
description of data at long wavelengths. The main goal to develop this model was to present a
global approach to interstellar dust modeling based on laboratory
measurements, and taking into account dust evolution in the different
gas phases. In addition, this model was configured with the use
of Galactic data only, which is probably not optimal
when analyzing dust, and in particular the smallest dust particles in
the LMC. 
DBP90 and MC11 are the two best models in this present analysis to
reproduce the PAH bands (with a maximal residual of -14.7$\%$ at 3.6
$\mic$) but DBP90 does not satisfactorily reproduce long-wavelength data. However, this model is the most simple of all the
models presented here, and has less free parameters compared to the other
models. 

\section{Submillimeter flattening}
\label{sec_flattening}
During recent years, many studies have evidenced a flattening of
the spectra in the submm--mm, based on observational data \citep[see
  for instance][]{Israel10, Bot10, Galliano11, Paradis12, Gordon14}, but also based on laboratory data from grain analogs \citep{Demyk17}. Whereas several possibilities have been
proposed, its origin is still not understood. All the models used here
that incorporate a constant emissivity spectral index in the submm--mm
domain, that is, all models except AJ13, reveal
important absolute variations between residuals at 250 $\mic$ and 500
$\mic$ in the HI and H$\alpha$ phases. This result indicates that the
emission spectra derived from the models are systematically steeper compared to the data. 
The absolute variations of the residuals between the 250
and 500 $\mic$ are around 7$\%$ to 17.5$\%$ in the HI phase, depending on the model. In AJ13, the large carbonaceous
grains allow to flatten the spectra in the submm thanks to the low-emissivity spectral index of this component taken close to one. This is the
reason why this model is able to reproduce the data with minimal
residual at long wavelengths. As a consequence, the flattening of the spectra
in the submm-mm cannot be evidenced by analyzing the residuals with AJ13. In the H$\alpha$ phase, residuals between 250 and 500
$\mic$ also significantly vary in absolute value with a minimum of
$\simeq$11.3$\%$ with MC11 and a maximum of
$\simeq$17.6$\%$ with DBP90. This is the first time
that evidence has been found of a submm flattening in the ionized gas. 
However, in the CO phase the absolute variations are significantly
lower, between 0.43$\%$ and 6.7$\%$, and within the 7$\%$  calibration
uncertainties. These results indicate that
statistically dust emission spectra in the molecular phase are steeper
than in the HI and H$\alpha$ phases. 
Some studies have highlighted the decrease of the dust emissivity
spectral index in the diffuse ISM \citep{Paradis12,Planck14XVII}, and we show here for
the first time a change in the emissivity spectral index of the
various gas phases associated with the molecular clouds. Our results indicate
similar behaviors of the dust emission at long wavelengths in the HI
and H$\alpha$ phases.

We investigated the 500 $\mic$ residuals as a function of the
relative abundances of the different dust components and as a function
of the ISRF intensity (see Fig. \ref{fig_excess_ab}). The comparison of the
results obtained with the different dust models does not allow
any clear trends or correlations, nor anti-correlations, to be identified between the
parameters. Indeed, some trends are clearly model dependent. It is therefore not possible from these findings to evoke any possible links
between the 500 $\mic$ residual and the intensity of the radiation
field or the dust abundances. 
\begin{figure*}
\begin{center}
\includegraphics[width=17cm]{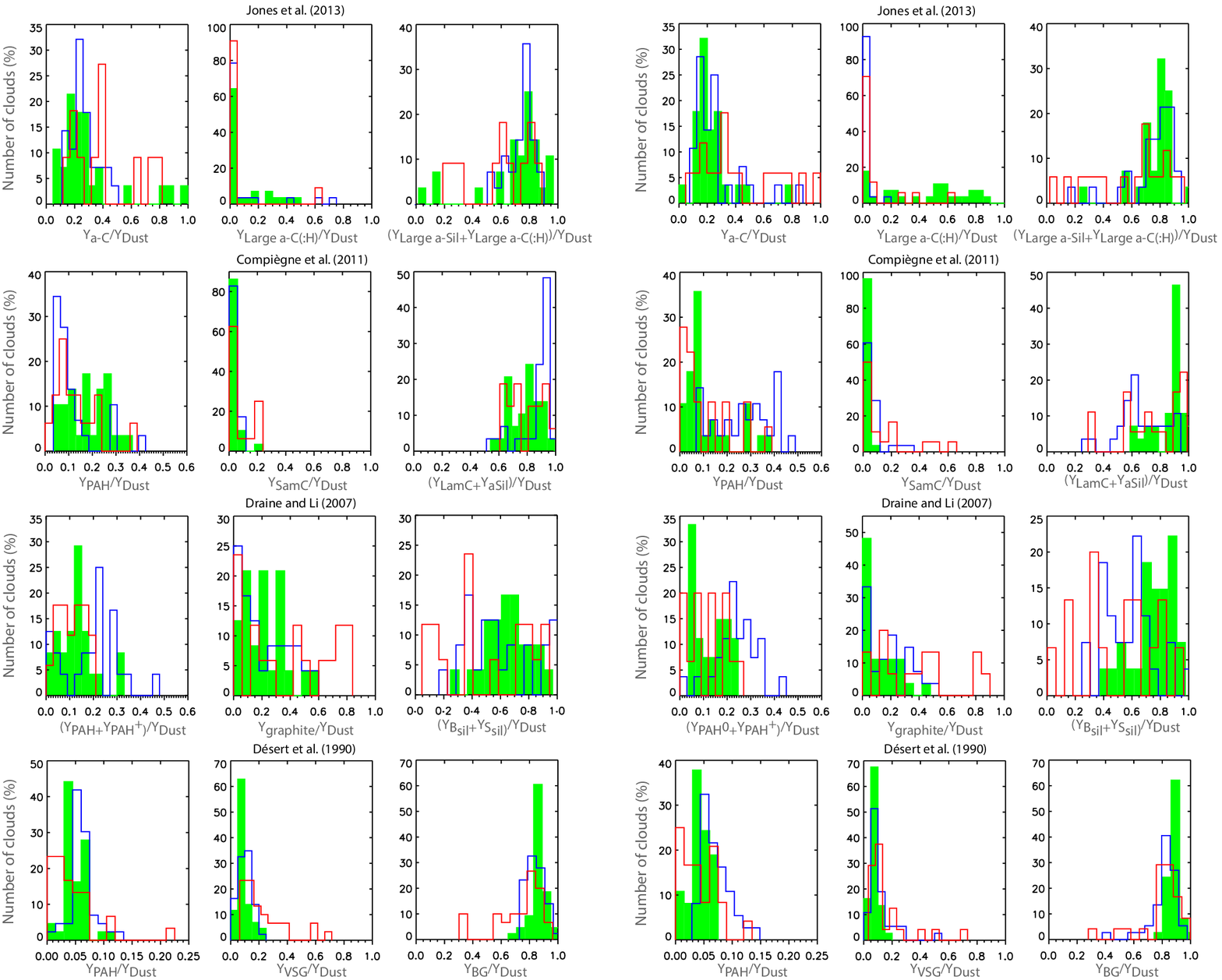}
\caption{Histograms of the relative abundances of the dust components
  for clouds with similar dust temperatures in the atomic and
  molecular phases (case 1 as defined in Sect.
  \ref{sec_temperatures}) on the left, and for clouds with
  significantly colder dust temperatures in the molecular than in the
  atomic phases (case 2) on the right. The atomic, molecular, and ionized phases are shown in
  green, blue, and red, respectively. We clearly evidence a
    positive shift in
  the distribution of the PAH relative abundance in the molecular
  phase for clouds in case 2 as compared to clouds in case 1. The VSG relative abundance seems to be increased in the
  ionized phase in both cases. \label{fig_hist_coldco}}
\end{center}
\end{figure*}
\section{Dust properties}
\label{sec_dustprop}
\subsection{Intensity of the radiation field and dust temperature}
\label{sec_temperatures}
All models show a statistical decrease of the
intensity of the ISRF in the molecular phase as compared to the atomic
one (see Fig. \ref{fig_param_AJ13}). In the case of MC11, the
decrease of $X\rm _{ISRF}$ in the molecular phase as compared to the
atomic one is not as visible as with the other models, but it exists.
Table \ref{table_xisrf} summarizes the results of Gaussian fits performed on
the histograms of the atomic and molecular phases. The distribution
for the ionized phase cannot be reproduced by a Gaussian fit. According to the model, the radiation field has a lower intensity in the molecular than in the atomic phase in 61$\%$ (DL07), 63$\%$ (MC11), 83$\%$ (DBP90) or 87$\%$ (AJ13)
of the clouds.  
Conversely, the ionized phase is statistically warmer
than the other phases, as visible in the histograms.

The DustEM software considers a size distribution for the different
grain component and therefore
BGs do not have a single equilibrium temperature. However, we
can compute an ``approximate'' single dust temperature. Indeed, along
the line of sight (LOS) we have:
\begin{equation}
  I_{\nu} \propto \int _{\rm LOS} \epsilon N_{\rm H} Y_{\rm Dust} X_{\rm
    ISRF},
  \end{equation}
 where $\epsilon$  is the dust emissivity, $N_{\rm H} $ the gas
column density, $Y_{\rm Dust}$ the dust abundance, and $X_{\rm
  ISRF}$ the intensity of the radiation field. Assuming a gray-body emission with a power-law
emissivity, and in the Rayleigh-Jeans limit where $B_\nu(T)
\propto 1/\lambda^4$, we obtain the following relation
\citep{Paradis11a}: 
\begin{equation}
T{\rm _d^{LMC}}=\left ( X{\rm _{ISRF}^{LMC}} \right
)^{1/(4+{\rm\beta)}} \times T{\rm _d^{\odot}}
,\end{equation}
where $\beta$ corresponds to the emissivity spectral index. We assume a dust
temperature in the solar neighborhood ($T_d^{\odot}$) of 17.5 K
\citep[using $\beta=2$,][]{Boulanger96}. The emissivity spectral index
of our models is fixed and close to a value of two, except with AJ13 when the large grain component is dominated by
carbon grains. However, when looking at Fig.
\ref{fig_param_AJ13}, histograms of $Y_{\rm Large\,a-C(:H)}$/$Y_{\rm
  {Dust}}$ and $Y_{\rm Large\,a-Sil+Large\,a-C(:H)}$/$Y_{\rm
  {Dust}}$ highlight the existence of
a significant fraction of clouds with large silicate grains
dominating the FIR--submm domain. 
We know that $\beta$ could vary with
wavelength and in particular in the submm--mm domain \citep{Meny07,
  Paradis11b, Demyk17}. However, this behavior is not systematic and has not
been observed in some galaxies \citep[see for instance][]{Galametz11, Aniano12}. However, around
the peak of BG emission, a range which is crucial to deduce the dust
temperature, $\beta$ is usually consistent with a value of two. 
Adopting a lower spectral index will better fit the dust emission at
long wavelength but can bias the determination of the dust temperature.
Therefore assuming a spectral index of two is definitely reasonable. 
The determination of a single dust temperature in each phase of the gas is useful to distinguish
different cases of clouds. We then define four cases of clouds with
threshold values that are slightly adjusted from one model to the
other to account for almost as many clouds in case 1 as in case 2:\\
- Case 1: clouds with similar temperatures in the atomic and molecular phases
(with a difference of temperatures $|T{\rm _d^{HI}}-T{\rm _d^{CO}}|$ $<
T{\rm _{tresh,low}}$, with $T\rm_{tresh,low}=$ 1 K for AJ13 and DBP90, 0.5 K
for MC11, and 0.6 K for DL07). \\
- Case 2: clouds with significantly colder dust in the molecular phase than
in the atomic one (with a difference of
temperatures $T{\rm _d^{HI}-T_d^{CO}}>T{\rm _{tresh,high}}$, with
$T{\rm _{tresh,high}}$= 4 K for AJ13, and 3 K for the other models). \\
- Case 3: clouds with colder dust in the molecular phase than
in the atomic one (with
$T{\rm _{tresh,low}}<T{\rm _d^{HI}}-T{\rm _d^{CO}}<T{\rm _{tresh,high}}$), excluding clouds in
case 2.\\
- Case 4: clouds with warmer dust in the molecular phase than
in the atomic one (with a difference of
temperatures $T{\rm _d^{CO}}-T{\rm _d^{HI}}>T{\rm _{tresh,low}}$).

We obtain:
\begin{itemize}
\item{With AJ13: 16$\%$ of clouds in cases 1 and 2, 62$\%$ in case 3, 6$\%$ in case 4} 
\item{With MC11: 19$\%$ of clouds in both cases 1 and 2, 30$\%$ in case 3, 32$\%$ in case 4} 
\item{With DL07: 15$\%$ of clouds in case 1, 17$\%$ in
    case 2, 33$\%$ in case 3, 35$\%$ in case 4 }
\item{With DBP90: 24$\%$ of clouds in case 1, 21$\%$ in
    case 2, 45$\%$ in case 3, 10$\%$ in case 4} 
\end{itemize}
This selection of clouds based on the dust temperature (specifically cases 1 and
2) is used in Sect. \ref{sec_ab} to analyze the trend of dust
abundances in clouds depending on their dust temperatures in the
atomic and molecular phases. 


\subsection{Relative dust abundances}
\label{sec_ab}
Figure \ref{fig_param_AJ13} shows histograms of the
different dust component abundances for each gas phase derived from
the models. We can identify an enhancement of the relative abundance of 
the PAH component (not directly observable with AJ13 since the a-C
component is a mixture of small grains and PAHs and is described by a
single dust component) in the molecular phase, whatever the model used. This can
be visible with a positive shift of the entire histogram or part of it in the molecular phase. 
The ionized phase tends to show the opposite
behavior, that is, a decrease of the PAH relative abundance. 

The dust component accounting for the MIR emission (hereafter VSG component) is described by the SamC, graphite, and VSG components
according to the models. Histograms in the ionized phase are more
extended than histograms in the other phases and mainly show an
increase of this
VSG component in the ionized gas. 

The histogram of the BG relative abundance is directly linked to the
behaviors of the other dust components. Since the BG component accounts
for the most important part of the total dust mass (as clearly
seen in the histograms), an increase in the
PAH or VGS relative abundance will automatically result in a decrease of the
BG relative abundance.
For a comparison, Fig.
\ref{fig_param_AJ13_nhx} of the Appendix presents the histograms of the relative dust
abundances when including the dark gas component in the analysis. The
previous results remain unchanged. 

To go further in our analysis,  Fig. \ref{fig_hist_coldco} compares the dust component
abundances in clouds with similar dust temperatures in the atomic and
molecular phases (case 1 as defined in Sect. \ref{sec_temperatures}),
and in clouds with significant cold dust in the molecular phase (case
2). For clouds of case 1, one trend can be seen: an increase of the VSG relative abundance in the ionized
phase. Concerning the PAHs for clouds in case 1, results cannot been interpreted since
they are different depending on the model. When looking at clouds in
case 2, we observe two significant behaviors: an increase of the PAH relative abundance in the
molecular phase and an increase of the relative abundances of the VSG
component in the ionized phase. The apparent decrease of the BG relative
abundance in the molecular and ionized phase (except with AJ13) only
results from the increase of the PAH and VGS relative abundances in these phases. The
enhancement of the VSG relative abundance in the ionized
phase of clouds is observed in both cases 1 and 2, and therefore does not seem to be related to
the dust temperature in the gas phases. The inclusion of the dark gas
component in the analysis does not change any of these results (see
Fig. \ref{fig_hist_coldco_nhx}).


\subsection{NIR continuum}
\label{sec_nircont}
The origin of the NIR continuum is unclear but this ``component'' is required in most
cases to explain the dust emission spectra in the NIR, where the PAH
component alone fails. Table \ref{table_nircont} presents
the percentage of SEDs that require a NIR continuum, whatever its
intensity. AJ13 shows low percentages in the HI
and CO phases but is also not able to fully reproduce the NIR
observations. 
The distributions of the intensity of this component over the total
dust abundance are presented in Fig. \ref{fig_param_AJ13}. The distribution
in the ionized phase is more extended than the ones in the atomic and
molecular phases. High values of the
NIR continuum are required in the ionized phase to reproduce the SEDs in the NIR, whereas the
PAH relative abundance is also reduced in this phase (see Sect.
\ref{sec_ab}). If the PAHs were the continuum carriers then we would
expect to see variations of these two quantities going in the same
directions, and not in the opposite way as observed in our
study. \citet{Flagey06} analyzed the Galactic diffuse ISM in the near-to-mid IR domain and concluded that neither
scattered light nor PAH or VSG fluorescence could be responsible of
this continuum. 

\begin{figure*}
\begin{center}
\includegraphics[width=14cm]{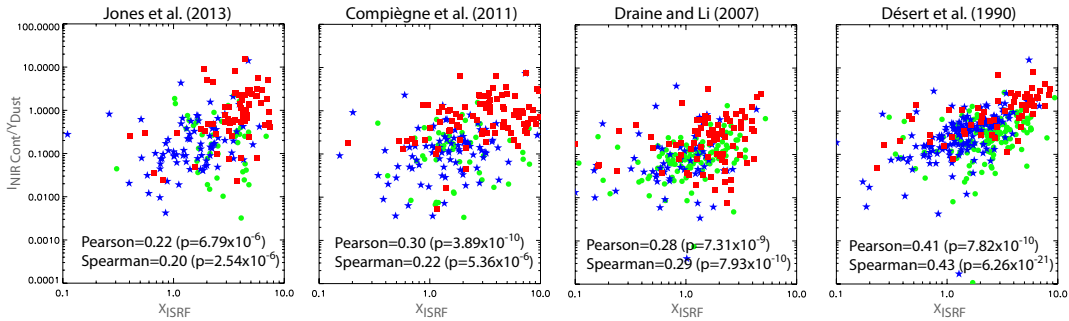}
\caption{Intensity of the interstellar radiation field as a function
  of the intensity of the NIR continuum normalized to the total dust
  abundance, for the different dust models. The atomic, molecular, and ionized phases are shown in
  green, blue, and red, respectively. The Pearson and Spearman
  correlation coefficients are given in each panel. \label{fig_cont_xisrf}}
\end{center}
\end{figure*}

We looked for possible correlations between the continuum
and the relative abundance of the different dust components, and no
direct link was found. However, the NIR continuum normalized to the
total dust abundance seems to be connected to the intensity of the
radiation field, as visible in Fig.
\ref{fig_cont_xisrf}, and this is confirmed by correlation coefficients such
as Pearson (going from 0.22 to 0.41 depending on the model) and
Spearman (going from 0.20 to 0.43). The p-values associated to
  each of the correlations are very low (see Fig.
  \ref{fig_cont_xisrf}) indicating that the correlation is likely to
be real. The origin of this continuum is further
discussed in Sect. \ref{sec_discuss_nircont}.

\begin{table}
\caption{Percentage of SEDs for which the presence of a 
  NIR continuum is required, as a function of the dust emission models.\label{table_nircont}}
\begin{center}
\begin{tabular}{lcccc}
\hline
   & AJ13 & MC11 & DL07 & DBP90 \\
\hline
\hline
HI       &19 &34& 69&82 \\
\hline
 CO      &29 &58&37&89 \\
 \hline
 H$\alpha$ & 65&78&81&68 \\
\hline
\end{tabular}
\end{center}
\end{table}
\begin{table*}
\caption{Summary of the method used in this work and
  \citet{Paradis11a}, and the different tests we performed to allow the
  comparison of these two studies. This table does not show all the results of
  this present analysis, but only results that can be directly
  comparable with \citet{Paradis11a}. Columns 2 and 3 describe the angular resolution,
  the presence or not of the ionized and dark gas phases (in addition
  to the atomic and molecular phases), the
  wavelength range, the model(s) used, and the results of these two
  studies in terms of dust temperature and PAH relative abundance in
  the atomic and molecular phases. Column 4 describes the different
  tests we applied, indicating the angular resolution, the gas
  phases we considered in the analysis, and the model we used. Column 5 summarizes
  the results of the tests. \label{table_comp}}
\begin{center}
\begin{tabular}{lllll}
\hline
\hline
  & This work & \citet{Paradis11a} & Test & Results of the tests \\
\hline
  Resolution & 1$^{\prime}$ & 4$^{\prime}$ & 4$^{\prime}$, with HI/CO/H$\alpha$;&
                                                            $\rm T_d^{CO}$$\searrow$ and
                                                                             $\left
                                                                                  (\rm 
                                                                                  \frac{Y_{PAH}}{Y_{Dust}}
                                                                                  \right
                                                                                  )^{\rm
                                                                                  CO}$$\nearrow$
  \\
  & & & Model: DBP90& \\
  H$\alpha$ phase & Yes & No & 4$^{\prime}$, HI/CO (H$\alpha$ removed)
                                          & $\rm T_d^{CO}
                                                        \simeq
                                                                               T_d^{HI}$
                                            and $\rm \left ( \frac{Y_{PAH}}{Y_{Dust}} \right )^{CO}$$\nearrow$\\
  &&&Model: DBP90& \\
  Dark gas phase & No (yes) & Yes & 4$^{\prime}$, HI/CO+Dark gas
                                                     &
                                                                $\rm T_d^{CO}
                                                        \simeq
                                                                T_d^{HI}$
                                                           and $\rm \left ( \frac{Y_{PAH}}{Y_{Dust}}\right )^{CO}
                                            \simeq \left (
                                                                   \frac{Y_{PAH}}{Y_{Dust}}
                                                                   \right
                                                              )^{HI}$\\
  &&&Model: DBP90&  \\
  Wavelength range & 3.6-500 $\mic$ & 3.6-160 $\mic$ & -- & -- \\
  Modeling & AJ13, MC11  & DBP90 && \\
  & DL07, DBP90 && \\
  \hline
  Comparison of the results & $\rm T_d^{CO}$$\searrow$ &$\rm T_d^{CO}
                                      \simeq T_d^{HI}$&& \\
       &     $\rm \left ( \frac{Y_{PAH}}{Y_{Dust}} \right
         )^{CO}$$\nearrow$ & 
                                                      $\rm \left (
                                                                   \frac{Y_{PAH}}{Y_{BG}}
                                                                   \right
                                                                   )^{CO}
                                            \simeq \left (
                                                                   \frac{Y_{PAH}}{Y_{BG}}
                                                                   \right
                                                                   )^{HI}$
                                   & & \\
  \hline
  \hline
  \end{tabular}
\end{center}
\end{table*}

\section{Discussion}
\label{sec_discussion}
\citet{Paradis11b} made the first study of dust in the ionized medium of
the LMC at 4$^{\prime}$ angular resolution. This analysis was done at the Galaxy
scale in three regimes of the ionized gas: diffuse, typical and bright
HII regions. This work highlighted several results: a decrease of the
PAH relative abundance in the ionized phase, as well as an increase of
the VSG relative abundance in the ionized phase of bright HII regions
compared to typical ones. This phase also
showed an enhancement of the NIR continuum. On the other hand,
the molecular phase seems to allow the survival of the PAHs. This
study was done using the updated DBP90 model. We obtain the same results in this present analysis, at small
scale in the molecular clouds, with a larger spectral coverage,
a better angular resolution, and with the use of four distinct dust
models to avoid conclusions that could be model-dependant. We also
showed a link between the NIR continuum and the intensity of the
radiation field, as well as some variations linked to the BG component
(dust temperature and emissivity slope) according to the gas phases. On the contrary, \citet{Paradis11a} concluded that  statistically,
in  the  majority  of  the  molecular clouds of the LMC,  there  was  no  apparent  evolution
in  the  PAH  and  VSG  properties  between the  atomic and the
molecular phase at 4$^{\prime}$ angular resolution. The origin of this
disagreement has been explored and is discussed in the following section.

\subsection{Comparison with \citet{Paradis11a}}
\label{sec_comparison}
In \citet{Paradis11a}, we did not observe any changes in the dust
component properties between the
atomic and molecular phases of the molecular clouds. However, in this
present work the method
is different for four reasons: \\
- (1): the angular resolution is higher, due to the 1$^{\prime}$ resolution of the Mopra instead of the
4$^{\prime}$ NANTEN data; \\
- (2): the ionized gas
phase associated with the molecular clouds is included in
this analysis and was not in \citet{Paradis11a}; \\
- (3): the dark gas is not taken into account \citep[as opposed to][]{Paradis11a}, but we
have checked the consistency of the results if it is included in the analysis; \\
-(4): the wavelength range is extended from 160 $\mic$ to 500 $\mic$.  \\
\citet{Paradis11a} performed the modeling with DBP90 only.
The main difference between the results
from \citet{Paradis11a} and those of the present study concerns the dust
temperature and the PAH relative abundance in the atomic and molecular
phases. Indeed, \citet{Paradis11a} did not observe any temperature
decrease or any increase of the PAH relative
abundance in the molecular phase. The comparison between the two studies is
summarized in Table \ref{table_comp}. Other results based on the analysis of the ionized phase or
on the submm flattening are not discussed here because these topics
were not addressed in \citet{Paradis11a}. 

To investigate the origin of theses discrepancies, we performed
  three tests (described below) with DBP90. Table \ref{table_comp} summarizes the
tests and their results.

We first probed the impact of the angular resolution on the results
  (point (1)). We reproduced this present analysis at
an angular resolution of 4$^{\prime}$ using NANTEN instead of
Mopra data. We also restricted our modeling to
the \citet{Desert90} model, as in \citet{Paradis11a}. The results indicate a
statistical decrease of the dust
temperature, as well as an increase of the PAH
relative abundance in the molecular phase as compared to the atomic
one. Results of this test are therefore identical to results of this
present work performed at 1$^{\prime}$ angular resolution. The lower
angular resolution in the previous analysis performed by \citet{Paradis11a}
does not explain the differences observed during this present study.  


Second, we probed the effect of the potential effect of the absence of the ionized phase
  in the IR decomposition at an angular resolution of 4$^{\prime}$  (point (2) combined to point (1)), as in \citet{Paradis11a}. We performed the IR
  decomposition using the atomic and molecular gas phases only. This time we do not observe a statistical
decrease of dust temperature in the molecular phase, but we still see an increase of the PAH
relative abundance in the clouds with significant colder dust in the
molecular than in the atomic phase. This test shows that the
  absence of the ionized phase in the analysis, as performed in
  \citet{Paradis11a}, can significantly affect the dust temperature
  derived in each phase. This potentially results from an impact on
  the shape of the SEDs in the FIR wavelength range only. However, this test still does not explain the disagreement in the PAH relative abundance.

Third, we probe the effect of the presence of the dark gas phase on the IR
decomposition, still working at an angular resolution of  4$^{\prime}$  (point
(3) combined with points (2) and (1)). The gas components therefore include the atomic,
molecular, and dark gas phases, as in \citet{Paradis11a}. In this
case, we still do not observe the statistical decrease of the dust
temperature in the molecular phase, as in the previous test, but this time the increase of the PAH relative
abundance in the molecular phase also disappears. We therefore recover
results obtained in \citet{Paradis11a}. It is mainly for this reason that we have
not performed further tests based only on the wavelength range.

These tests show that the results can be very sensitive to
  the gas components considered in the IR decomposition, depending on
the angular resolution. Indeed, the more pixels there are in the
decomposition, the more the shape of the SED will be unchanged, and the
more the results will be unmodified with the presence or not of an additional gas phase. In the present case, the
analysis is done at a small scale (molecular cloud scale). Furthermore, the study is sensitive to the number
of pixels per cloud and therefore to the angular resolution. For instance, at
1$^{\prime}$ resolution, we have seen that whether or not the dark
 gas is taken into account does not have any impact on the results. However, at
 4$^{\prime}$ resolution, ignoring the ionized phase has an
 impact on the dust temperature, and adding the dark gas phase
 significantly affects results on the PAH relative abundance.

 To conclude, the absence of the ionized gas in the
 dust decomposition in the study performed
 by \citet{Paradis11a}, combined with the inclusion of the dark gas (highly
 correlated to the atomic phase), at a lower angular resolution (4$^{\prime}$), explains the
 differences observed between this latter study and the present one.


\subsection{Grain aggregation}
In this present work based on the analysis of molecular clouds at an angular resolution of 1$^{\prime}$
, we
reveal some dust evolution between the atomic and molecular phases:\\
- most of the molecular clouds show colder dust in the molecular phase
than in the atomic
one; \\
- statistically, dust emission spectra in the molecular phase
are steeper than in the atomic phase; \\
- clouds from case 2 (clouds with much colder dust in the molecular than in the atomic phase) point out an increase of
the PAH relative abundance. \\
Furthermore, the ionized phase shows a similar submm flattening to that in
the atomic phase, compared to the molecular phase.

Grain aggregation is expected to occur in cold molecular clouds. Big grains could
aggregate with one another due to the presence of ice mantles on their
surfaces \citep{Stepnik03, Kohler11, Kohler12} but could also aggregate VSGs. Here we do not see any signs of
VSG disappearance as evidenced in \citet{Ysard13} and \citet{Tibbs16}. \citet{Tibbs16} observed a depletion of VSGs ($\it a$$<10$ nm) in a sample
of Galactic cold cores using 1 cm CARMA observations. They used
spinning dust emission observations to constrain the abundance of
VSGs and attributed their depletion to grain growth via accretion and
coagulation. However, this present study is based on several assumptions and
the main one is that the spinning dust hypothesis is the correct interpretation of the
anomalous microwave emission. In addition, the study by  \citet{Tibbs16} relies on the
analysis of 15 clouds, whereas the sample in our analysis is
significantly larger ($\simeq$ 170 clouds). However, our statistical
result could describe a general behavior and also hide some specific molecular clouds of the
LMC with a decrease of the VGS relative abundance in the molecular
phase. \citet{Stepnik03} and \citet{Ysard13} observed a
decrease in the abundance of small carbon grains  in a dense filament of the Taurus
molecular complex (L1506) using Herschel data. Their spatial scale
allowed them to reach significantly higher densities than what we can
probe in this present analysis due to the resolution and the
distance of the LMC. Therefore, the difference between the studies
could result from a difference in local density, from specific cases that
do not apply to most of the molecular clouds, or
from different dust properties of the LMC molecular clouds compared to those in our Galaxy.  

Big grain coagulation from the diffuse to the dense medium could induce a change in the dust
emissivity spectral index between the two phases \citep{Paradis09},
evidenced here by steeper dust emission spectra at submm wavelengths in the molecular phase as compared to the atomic phase. 

\citet{Jones13} proposed a scenario of dust evolution with density,
based on the formation of a second mantle of more
or less aliphatic matter accreted on top of the aromatic mantle. This model predicts higher spectral index in molecular
clouds (because of small particle accretion and a-C:H mantle
accretion) than in the diffuse ISM (large homogenous
a-C grains). However, because a-C(:H) materials are significantly
less emissive than a-C grains, their effects on the
FIR--submm emissivity may not be observable. In
addition to the second mantle, an ice mantle can
be formed at the surface of the grains, favoring
grain coagulation. Such a scenario may explain the observed steeper spectra in dense phases relative to the atomic and ionized phases (the
second mantle being destroyed by high-intensity
radiation fields). 

In the framework of the TLS model \citep{Meny07},
the change in the emissivity spectral index could be the result of variations in the degree
of amorphization of the grains, that is, the disorder at atomic
and/or nanometer scales \citep{Paradis12}. Concretely,
to explain the FIR-to-submm flattening in the diffuse and ionized phase, the structure
of the dust should be different in the atomic and the ionized phase as
compared to the molecular  phase; it should be more altered, containing more
defects or voids for example in the atomic/ionized phase than in the
molecular phase. Such an alteration could be the result of dust processing by low-energy
cosmic rays. Strong shocks due to supernova explosions could be another alternative to alter dust in the atomic
and ionized phases. Moreover, shocks could
destroy some of the largest grains and increase the amount of very
small grains, as observed in the ionized phase. Supernova explosions would
also contribute to the increase of the dust temperature in the
ionized phase.

These two previous scenarios coming from different dust models (AJ13
and TLS) could explain the fact that no link between the
submm flattening and the small dust component abundances has been
found. 

\subsection{Grain formation/destruction}
Polycyclic aromatic hydrocarbons (PAHs) seem to be destroyed in the ionized phase, whereas their relative
abundance is enhanced in the cold molecular phase, as observed in
\citet{Paradis11b}. The increase of the radiation field in the ionized phase
might be responsible for their destruction. \citet{Sandstrom10} also
pointed out a low-mass fraction of PAHs in the diffuse regions of the
Small Magellanic Cloud compared to a higher fraction in dense
regions. Their study is based on the \citet{Draine07} model.  It
appears therefore that both Magellanic Clouds show an enhancement of the PAH
relative abundance in the dense phases. \citet{Sandstrom12} propose a scenario in which PAHs
are formed in regions of dense gas, with a smaller average size, and are
more neutral than in galaxies with higher metallicity. This scenario proposes that they then go on to emerge in the diffuse
ISM where a fraction of them could be destroyed under typical ISM conditions, since UV
photons, shocks, cosmic rays, or hot gas destroy small
PAHs more efficiently than large PAHs. Our analysis
indicates that the PAH bands in all gas phases are best reproduced by
MC11 and DBP90, with full neutral
PAHs, but with similar dust size distribution to that in our Galaxy. Our
study does not seem to favor the hypothesis of smaller PAHs in the
LMC, as compared to the Milky Way for instance. However, our present
analysis could also indicate the destruction of PAHs by the higher-intensity radiation field, as PAHs do not disappear in
the atomic gas which has typical ISM conditions. Moreover, PAHs could
be formed in clouds with significantly colder dust in the molecular
phase compared to the atomic phase.

The relative abundance of VSGs is enhanced in the ionized phase, but
does not seem to show any link to the nature of the clouds (quiescent or with star formation activity), since their
abundances increase in the ionized phase whatever the intensity of the radiation
field. Therefore, VSGs could mainly be formed
in HII regions rather than in molecular clouds. Since we do not
see any signs of VSG coagulation onto BGs, we cannot favor
the hypothesis according to which VSGs are released from the surface of
BGs to explain the increase of their abundance in the ionized
phase. However, another possibility could be the destruction of larger BGs in shocks.

\subsection{Near-infrared continuum}
\label{sec_discuss_nircont}
In Sect. \ref{sec_nircont} we showed a correlation between
the intensity of the NIR continuum and the intensity of the
interstellar radiation field. However, no link was found between
the continuum and the relative abundances of the different dust
components. This continuum was first observed in reflection nebulae
\citep{Sellgren83}, and then in the Galactic ISM
\citep{Flagey06}, but also in high-redshift galaxies
\citep{Mentuch09}. The latter authors detected this continuum in a
sample of 88 galaxies from the Gemini Deep Survey. They concluded
that circumstellar disks of young massive objects are the most likely candidates to explain
the 2–5 $\mic$ excess. 
At the Galaxy scale where the total IR luminosity is dominated by
massive stars, this hypothesis could be plausible but at small scales
where we discern the IR emission coming from the different gas
phases, it would be difficult to explain the presence of the
NIR continuum in the atomic phase and in the cold molecular phase. In
addition, \citet{Lu04} showed that the average 3 to 10 $\mic$ spectral shape
of the ISM emission does not vary significantly from diffuse sources to star-forming
regions. 

\citet{Duley09} investigated the origin of the NIR continuum in the
ISM and suggested that carbon molecules and especially
dimers of dehydrogenated carbon molecules with less than
28 atoms could be heated to high temperatures ($\simeq$ 1500 K) by
absorption of UV photons. This could explain the link between the
intensity of the NIR continuum and the intensity of the interstellar
radiation field. Furthermore, this assumption could explain the fact
that the continuum appears to unrelated to the relative
abundances of the dust components used in the models. 

\section{Conclusions}
\label{sec_cl}
We performed a decomposition of the IR emission with the different gas phases (atomic, molecular, and
ionized phase, as traced by HI, CO, and H$\alpha$ emissions,
respectively) associated with the LMC molecular clouds. We tested the strength of our conclusions
by either taking into account the dark gas or not in the dust
 decomposition; the conclusions are not affected. The analysis was performed
at an angular resolution of  1$^{\prime}$  using Spitzer and Herschel data
for the IR emission, and ATCA/Parkes, Mopra, and SHASSA data for the
gas tracers. The SEDs resulting from the correlations were modeled
using four different dust models: \citet[][, AJ13]{Jones13},
\citet[][, MC11]{Compiegne11},\citet[][, DL07]{Draine07}, and an improved version of
\citet[][, DBP90]{Desert90}. Whereas all
models give similar fits (within a few percent) between 70 and 500
$\mic$, the 24 $\mic$ data require a change in the size distribution
of the a-C component in AJ13, as compared to the Galactic one. The main
differences in the model appear in the PAH bands, with the largest residuals
being seen in the case of AJ13. DL07 clearly
needs a mixture of neutral and ionized PAHs, whereas MC11 and DBP90 give satisfactory results
with fully neutral PAHs. This multi-modeling analysis is important to ensure the strength of each of
our results presented here.

 We found direct evidence for the evolution of 
 dust
 properties from the
 atomic to the molecular medium, that manifests through a decrease in the
 intensity of the radiation field in the molecular phase affecting the
 dust temperature. Furthermore, we
 observe a change in the slope of the emission spectrum in the
 molecular phase, and in particular a steeper spectrum in the FIR to submm
 range. These results could be the consequence of BG coagulation in
 the dense phase. In addition, the cold molecular
 phase tends to favor the increase of the PAH relative abundance. However,  dust evolution was not found
in a previous study made by \citet{Paradis11a}. We
 investigated the difference in the results between this latter study and the present one and concluded that the absence of the ionized gas phase in the
 dust decomposition, combined with the inclusion of the dark gas (highly
 correlated to the atomic phase) in the previous study, significantly affects the shape of the
 SEDs at small scale. The reason is that at small scale and at 4$^{\prime}$ angular resolution the number of
 pixels is significantly lower than at 1$^{\prime}$ angular
 resolution, which has an impact on the SED and the derived parameters.

For simplicity we name VSG as the dust component responsible of the
MIR emission (originating from DBP90). This component corresponds
to graphites in DL07 and small amorphous carbons
in MC11. In AJ13, the amorphous carbon ($a-C$) component includes PAHs and
does not allow us to analyze each component separately.
Our analysis does not show any signs of 
 disappearance of VSG that could result from
 coagulation on the BGs. However, this process could occur only
 at very high density, something that we cannot probe in this analysis, or could apply
 to specific cases and not to the overall molecular clouds. Another
 explanation could be 
 that molecular clouds in the LMC have different dust properties as
 compared to those in the Milky Way. 
 
The use of various dust models confirms different properties in
the ionized gas: an increase of the dust temperature, an increase of the VSG relative
abundance and the intensity of the NIR-continuum, and a decrease of
the PAH relative abundance. We attribute these differences in the PAH
and VSG abundances to different locations of grain
formation: VSGs could be formed in HII regions rather than in
molecular clouds, whereas PAHs could be formed in the molecular phase
of cold and dense clouds. The NIR-continuum does not
seem to originate from any single dust component included in the dust
models. We also showed that a correlation exists between the relative intensity
of the continuum and the
intensity of the radiation field. We therefore do not favor the
hypothesis of PAH  molecules as carriers of the NIR continuum, nor
circumstellar disks as possible candidates. We do not reject the
possibility that another small dust component could explain this continuum, such as small carbon
molecules heated to high temperatures as proposed by
\citet{Duley09}. 

\begin{acknowledgements}
We are grateful to the referee for his/her careful reading and for all
the suggestions which helped to improve the quality of the manuscript. We acknowledge the  use  of  the  DustEM  package. MJ acknowledges the
support of the Academy of Finland Grant No. 285769. 
\end{acknowledgements}

\begin{appendix}
  \section{Results when including the dark gas in the dust decomposition}
  \begin{figure*}
\begin{center}
\includegraphics[width=16cm]{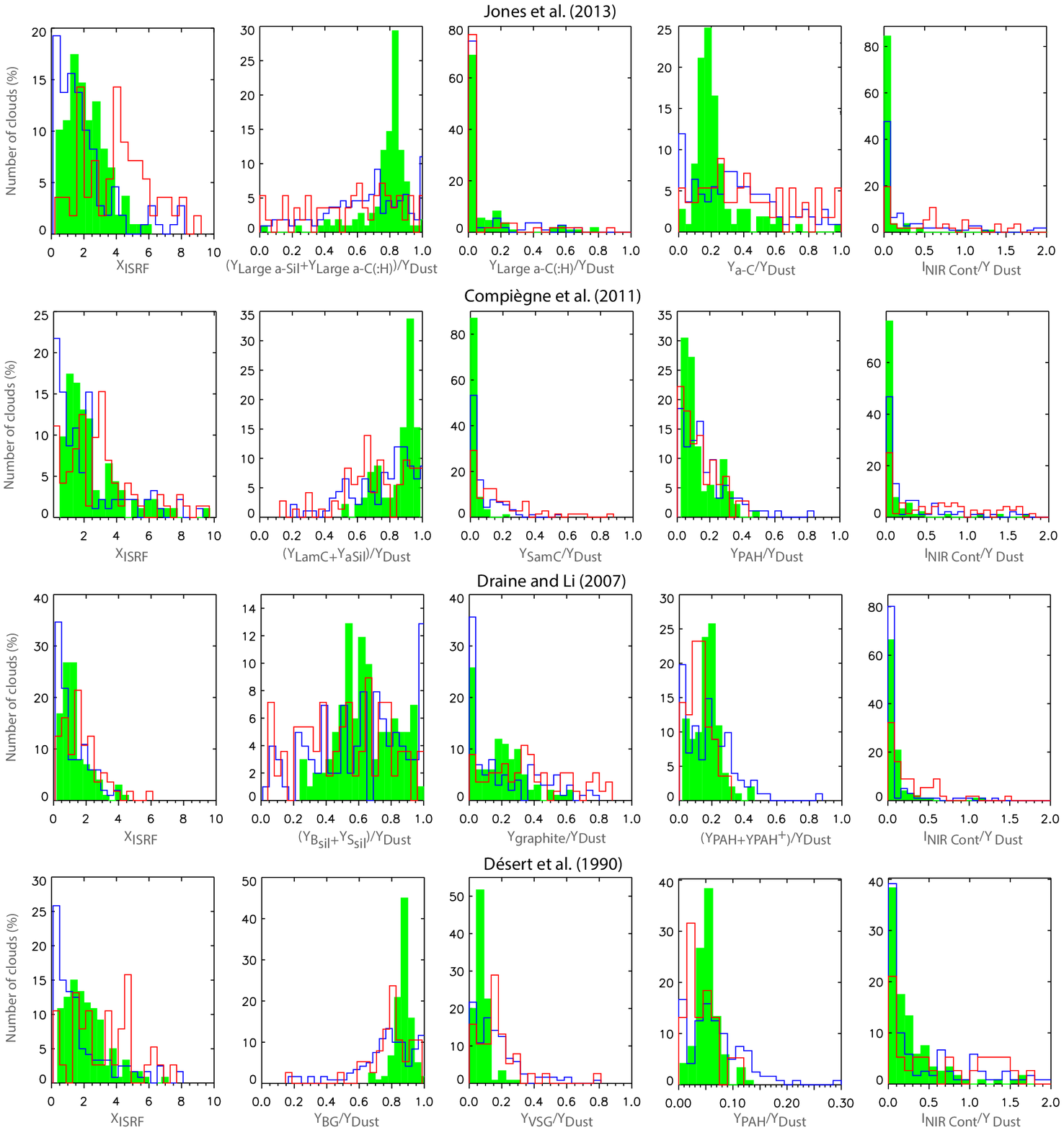}
\caption{Histograms of the parameters derived for each dust model (relative dust abundances, intensity of the radiation field, and relative intensity of the NIR
    continuum, see Sect. 4.2) in
  the different gas phases (atomic in green, molecular in blue, and
  ionized in red), when including the dark gas phase in the
  decomposition of the dust emission (see
  Sect. \ref{sec_decomposition}). We observe the same trends as in
  Fig. \ref{fig_param_AJ13}: a decrease of the intensity of the radiation field in
the molecular phase and an increase in the ionized phase, as
compared to the atomic phase; an increase of the PAH relative abundance in the
molecular phase; an increase of the VSG relative abundance in the
ionized phase and in the molecular one (that was not observed in
Fig. \ref{fig_param_AJ13}); and a more extended distribution of the NIR continuum
in the ionized phase as compared to the other phases.\label{fig_param_AJ13_nhx}}
\end{center}
\end{figure*}

\begin{figure*}
\begin{center}
\includegraphics[width=17cm]{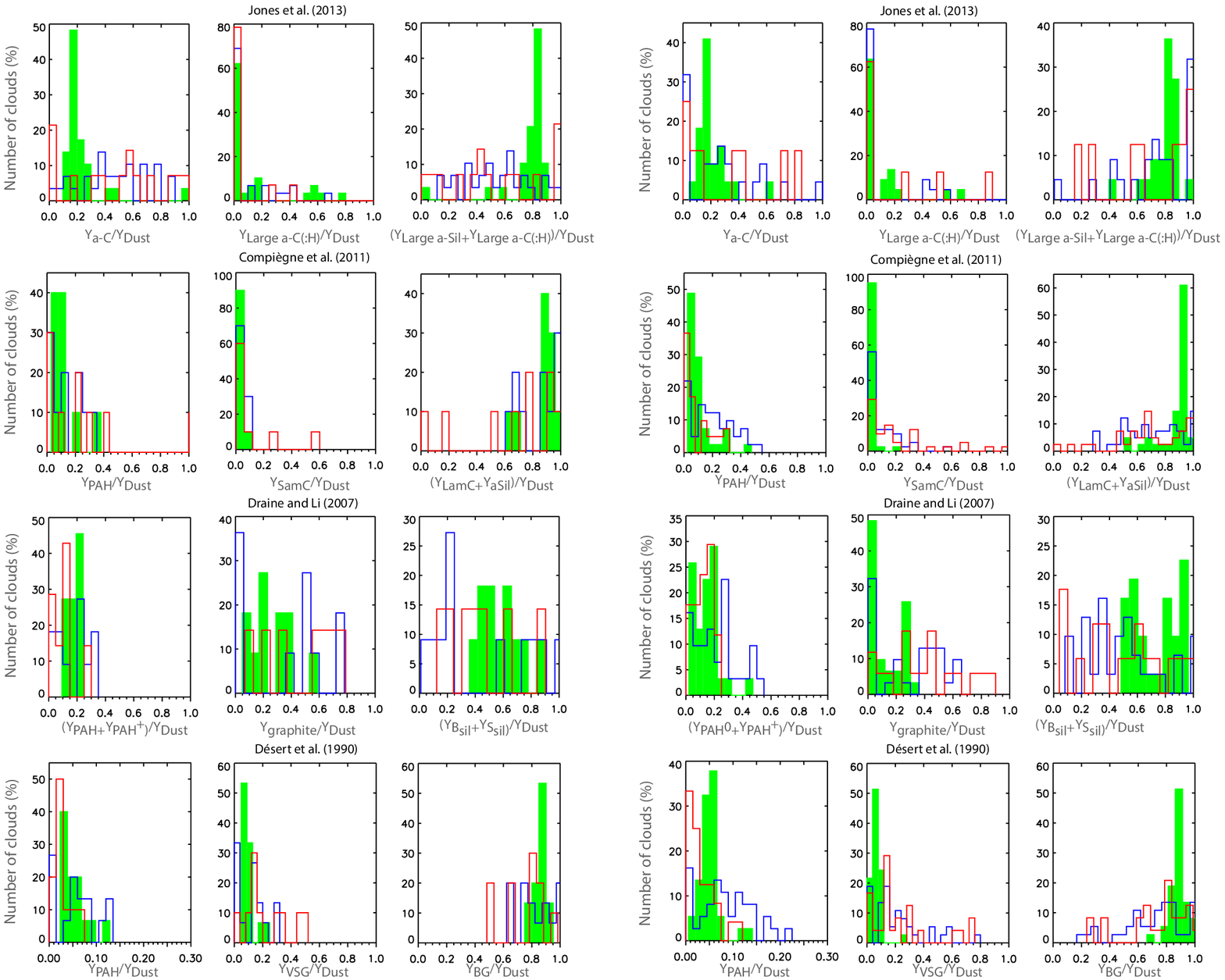}
\caption{Histograms of the relative abundances of the dust components
  for clouds with similar dust temperatures in the atomic and
  molecular phases (case 1 as defined in Sect.
  \ref{sec_temperatures}) on the left, and for clouds with
  significantly colder dust temperatures in the molecular than in the
  atomic phases (case 2) on the right. The atomic, molecular, and ionized phases are shown in
  green, blue, and red, respectively. The dark gas phase has been
  taken into account in the decomposition of the dust emission (see
  Sect. \ref{sec_decomposition}). As evidenced in Fig.
    \ref{fig_hist_coldco}, we notice an enhancement of the PAH relative abundance in the molecular
  phase (as compared to the atomic phase) for clouds in case 2 as
  compared to clouds in case 1. The VSG relative abundance is also increased in the
  ionized phase in both cases, as well as in the molecular phase (that
  was not observed in Fig. \ref{fig_hist_coldco}).\label{fig_hist_coldco_nhx}}
\end{center}
\end{figure*}

 \end{appendix}

\end{document}